\documentclass[aps,prl,reprint,superscriptaddress,nofootinbib,nobibnotes,floatfix]{revtex4-2}
\usepackage{graphicx}
\usepackage{bm}
\usepackage{amssymb}
\usepackage{amsbsy}
\usepackage{amsmath}
\usepackage{mathtools}
\usepackage{xcolor}
\usepackage{stmaryrd}
\usepackage{mathrsfs}
\usepackage[mathscr]{euscript}  
\usepackage{tikz}
\usetikzlibrary{shapes}
\usepackage{cancel}
\usepackage{booktabs}
\usepackage[normalem]{ulem}
\usepackage{etoolbox}
\usepackage{siunitx}
\usepackage{caption}
\usepackage{titlesec}
\usepackage{titlecaps}

\titleformat{\section}{\normalfont\bfseries}{\thesection}{1em}{}
\titleformat{\subsection}{\normalfont\bfseries}{\thesubsection}{1em}{}
\titleformat{\section}{\normalfont\bfseries\filcenter}{\thesection}{1em}{}
\titleformat{\subsection}{\normalfont\bfseries\filcenter}{\thesubsection}{1em}{}
\titleformat{\subsubsection}{\normalfont\bfseries\filcenter}{\thesubsubsection}{1em}{}



\makeatletter
\long\def\@makecaption#1#2{%
  \vskip\abovecaptionskip
  \sbox\@tempboxa{#1: #2}%
  \ifdim \wd\@tempboxa >\hsize
    {\setlength{\rightskip}{0pt}%
     \setlength{\leftskip}{0pt}%
     \parfillskip=0pt plus 1fil\relax
     \noindent #1: #2\par}%
  \else
    \global\@minipagefalse
    \hb@xt@\hsize{\hfil\box\@tempboxa\hfil}%
  \fi
  \vskip\belowcaptionskip}
\makeatother

\renewcommand{\figurename}{Fig.}

\begin{document}

\title{Active Particles Imprint Persistent Percolating Networks in Polymer Condensates}
\author{Ligesh Theeyancheri}
\email{ligeshbhaskar@gmail.com}
\affiliation{Department of Physics, Syracuse University, Syracuse, NY 13244, USA} 
\author{Jennifer L. Ross}
\email{jlross@syr.edu}
\affiliation{Department of Physics, Syracuse University, Syracuse, NY 13244, USA} 
\author{J. M. Schwarz}
\email{jmschw02@syr.edu}
\affiliation{Department of Physics, Syracuse University, Syracuse, NY 13244, USA} 
\affiliation{Indian Creek Farm, Ithaca, NY 14850 USA}

\begin{abstract}

\noindent Fluid condensates readily exchange components and reorganize, and in doing so typically erase structural history. Using simulations of sticker–spacer polymers in an active particle bath, we show that activity drives condensates from compact droplets into system-spanning percolated networks by enhancing interchain connectivity, suppressing intrachain collapse, and increasing topological constraints through interchain winding. The network persists after the active particles are removed, despite continued polymer exchange and contact turnover, revealing a fluid-like state with activity-induced topological imprinting. Hence, activity can write long-lived structural organization and memory into fluid condensates.
%\keywords
\end{abstract}

\maketitle

\begingroup
\renewcommand\thefootnote{}
\setlength{\footnotesep}{0.8\baselineskip}
\endgroup

 \noindent Conventional fluids continually rearrange and exchange constituents, progressively erasing structural organization. Solids, in contrast, readily retain structural history because their constituent network is effectively frozen. Biological systems often occupy an intriguing middle ground, maintaining persistent organization despite continual constituent turnover~\cite{Fletcher2010,Ranft2010,liu2021dynamic}. Inspired by such systems, can a fluid preserve its internal organization even as its microscopic constituents are continually replaced, and if so, where is that information stored?\\

\noindent Biomolecular condensates provide an ideal setting in which to investigate this question. Formed through liquid-liquid phase separation (LLPS), these membraneless organelles compartmentalize diverse cellular processes while remaining highly dynamic~\cite{banani2017, shin2017}. Driven by multivalent interactions, binding affinity, and patterning of molecular interactions, these condensates regulate processes from transcription and signaling to stress response and protein quality control~\cite{su2016phase, pei2025transcription, boeynaems2018, visser2024role, alberti2021,mittag2022, martin2020valence, farag2023}. Many of these biomolecular condensates often behave as seemingly conventional liquid-like droplets, undergoing coarsening, coalescence, and wetting, with rapid internal rearrangement and molecular exchange with the dilute phase as seen in RNA coacervates and nucleolar condensates~\cite{brangwynne2009, tang2015, feric2016}. Beyond simple droplet morphology, biomolecular condensates can adopt more complex morphologies typically driven by rheological changes to non-Newtonian fluids~\cite{tanaka2005, alshareedah2021, michieletto2022rheology, shen2023, das2023topological}. Recent force indentation experiments further reveal a rigidity transition within a multivalent protein condensate as protein binding strengths are modified~\cite{liao2025emergent, liao2026emergence}. In bacteria, PopZ condensates form a filamentous meshwork within rather spherical droplets, intriguingly~\cite{scholl2026}, highlighting the nontrivial aspects of condensates. \\

\noindent Biological condensates exist in an active cellular environment~\cite{sanfeliu2025mechano, berry2018physical, zwicker2025physics, wiegand2020drops, soggia2025mechanical}. Activity can potentially impact the organization of condensates.  Strikingly, a recent study demonstrated that an active bath of urease enzymes enhances Ubiquilin-2 droplet size and condensate fraction~\cite{ching2025}. In another study, active microtubule-kinesin flows in phase-separating polymer active-passive mixtures generate steady-state bicontinuous network morphologies inaccessible through equilibrium routes~\cite{gulati2026}. The former result can be understood in the context of an active bath composed of large excluded volume particles extending and corralling the protein polymers but subsequently becoming trapped within the droplets. The latter result can be understood in terms of an extensile active nematic generating turbulent-like flows inside the active fluid. These flows continuously stretch, fold, and deform the interface separating the active and passive phases to lead to a bicontinuous network. Remove the activity and remove the complex interfaces. Here, rather than asking only how activity modifies droplet size or shape, we ask a different question: can activity write structural information into a fluid condensate that persists after the activity itself has vanished?\\
\begin{figure*}[ht]
    \centering
    \includegraphics[width=0.9\textwidth]{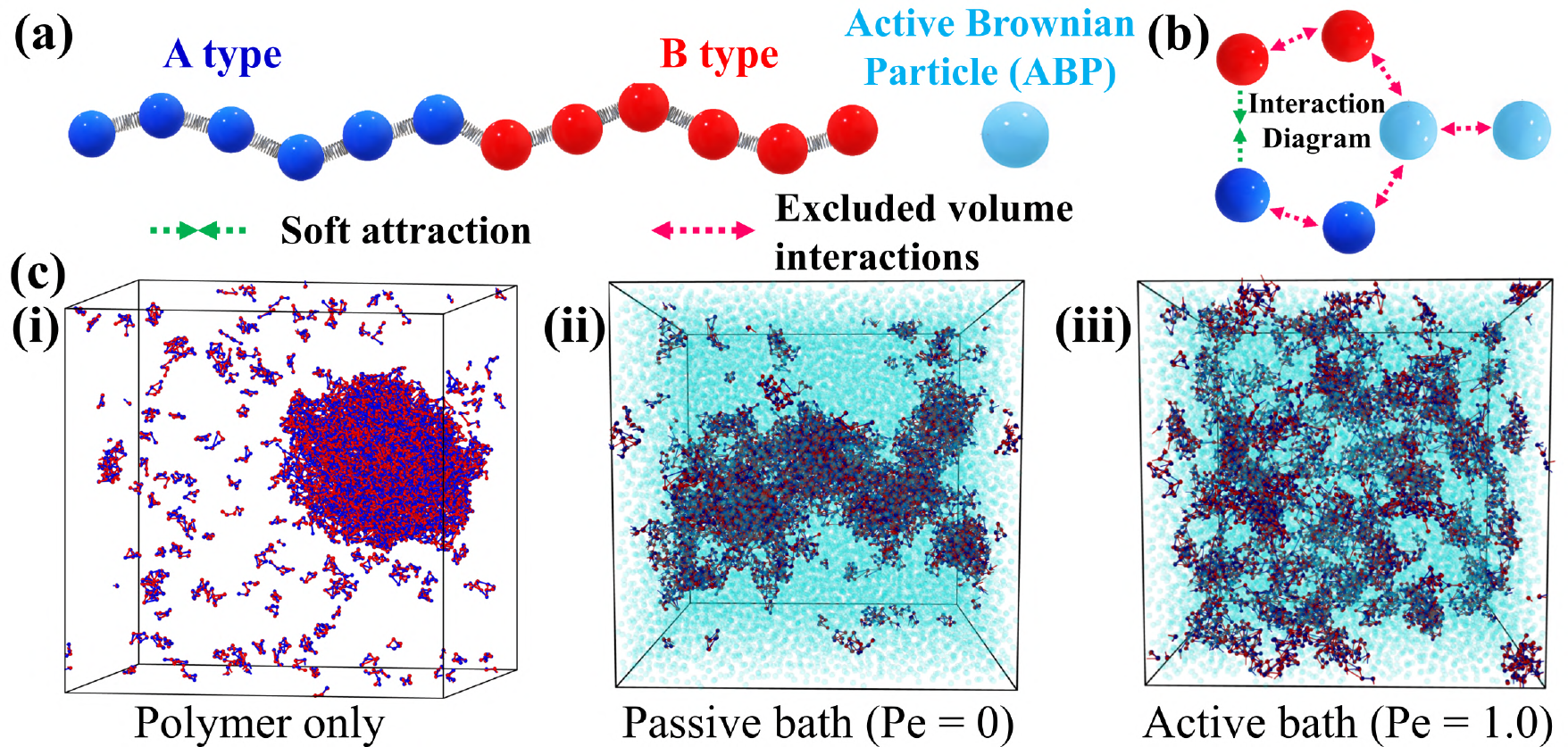}
    \caption{\textit{Activity-driven transition from polymer droplets to percolated networks.} (a) Sticker-spacer polymer model (A/B monomers) with active Brownian particles (ABPs) as the bath, (b) interaction scheme, and (c) representative steady states: (i) polymer-only (no bath), forming a compact droplet; (ii) passive bath ($\text{Pe}=0$), yielding compact dense clusters; (iii) active bath ($\text{Pe}=1.0$), producing a percolated network.}
    \label{fig:model}
\end{figure*}

\noindent We demonstrate that an active bath composed of excluded volume particles drives a structural transition in polymer condensates from compact equilibrium droplets to a nonequilibrium system-spanning percolated network that can persist even after the activity is removed. The memory is in the form of persistent topology. To demonstrate this phenomenon, we consider the established ``sticker--spacer'' polymer framework~\cite{zhang2024exchange, choi2020physical} to model the condensate, in which each chain of sequence $(A_n B_n)_m$ consists of A and B type monomers of diameter $\sigma_p = 2 \, nm$, connected by finitely extensible nonlinear elastic (FENE) springs as implicit spacers~\cite{kremer1990dynamics} (Fig.~\ref{fig:model}a), with $n$ and $m$ varied at fixed total chain length ($N_p = 12$) to isolate the effect of monomer patterning. Heterotypic A--B pairs interact via a soft attractive potential of strength $\epsilon_{AB} = 6 k_BT$, modeling specific, one-to-one binding, while same type monomer pairs interact through a purely repulsive Weeks--Chandler--Andersen (WCA) potential~\cite{weeks1971role} to prevent nonspecific aggregation. The active bath consists of $N_s$ spherical active Brownian particles (ABPs) of diameter $\sigma_s = 3\,nm$, which self-propel at constant speed along a body-fixed direction that undergoes rotational diffusion~\cite{bechinger2016active, theeyan2020translational, du2019study, volpe2014simulation}, and interact with all monomers through the WCA potential (Fig.~\ref{fig:model}b). The bath density is set by $\rho_{_{M}}=\frac{N_s}{N_p N}$ where $N$ is the total number of polymers, with $\rho_{_{M}}\in \{0.5, 1.0, 1.5, 2.0, 2.5\}$. The dynamics of all particles evolve under an overdamped Langevin equation:
\begin{equation}
\gamma \, \dot{\bm{r}}_i = - \nabla_i V(\bm{r}_{1},\bm{r}_{2},..., \bm{r}_{N}) + \bm{f}_{i}(t) + {F}_{a} \bm{\hat{n}}_i \label{eq:langevineq}
\end{equation}
where $\bm{r}_i$ is the position of the $i^{th}$ particle, $\gamma$ is the friction coefficient, $V(\bm{r}_1, \ldots, \bm{r}_N)$ includes all bonded and non-bonded interactions, and $\bm{f}_{i}(t)$ is a Gaussian thermal noise. The active force term $F_a \bm{\hat{n}}_i$ acts only on bath particles, where $F_a$ is the magnitude of the activity and $\bm{\hat{n}}_i$ is a unit orientation vector. Polymer monomers remain passive throughout. Activity is characterized by the dimensionless P\'{e}clet number $\text{Pe}$, which quantifies the ratio of active to thermal forces, defined as $Pe = \frac{F_a\sigma_s}{k_BT}$. The system contains $N = 1000$ polymers in a cubic box of side $a = 107.72 \, nm$ with periodic boundary conditions at $T = 300$K, corresponding to a polymer volume fraction of 0.04. Full details of the interaction potentials and simulation parameters are provided in the Supplemental Material (SM). In the absence of a bath, inter-monomer attractions drive equilibrium phase separation into a single compact droplet (Fig.~\ref{fig:model}c(i)) with a unimodal density profile centered on the droplet (Fig.~S1). \\ 

\noindent Steady-state configurations of the model are shown in Fig.~\ref{fig:model}c(ii, iii). Introducing a passive bath yields a dispersed large cluster with a few smaller ones in the dilute phase, where excluded-volume interactions suppress full coalescence while each cluster remains internally dense and compact (Fig.~\ref{fig:model}c(ii)). Conversely, making the bath active ($\text{Pe} = 1.0$) replaces these compact clusters with a system-spanning percolated network, with active bath particles filling the interstitial voids, demonstrating that activity drives a structural transition from isolated droplets to a percolated network of condensates (Fig.~\ref{fig:model}c(iii), Fig.~S2). To quantify the network connectivity and spatial extent, we compute the fraction of polymer chains in the largest connected component $\langle f_{\rm chain}^{\rm LCC} \rangle$ and the mean isotropic span $\langle \mathcal{S} \rangle$ of the LCC, both increasing monotonically with Pe (Fig.~S3), confirming that activity drives a progressively more connected and system-spanning network (see SM for details).\\ 
\begin{figure}[h!]
    \centering
    \includegraphics[width=0.48\textwidth]{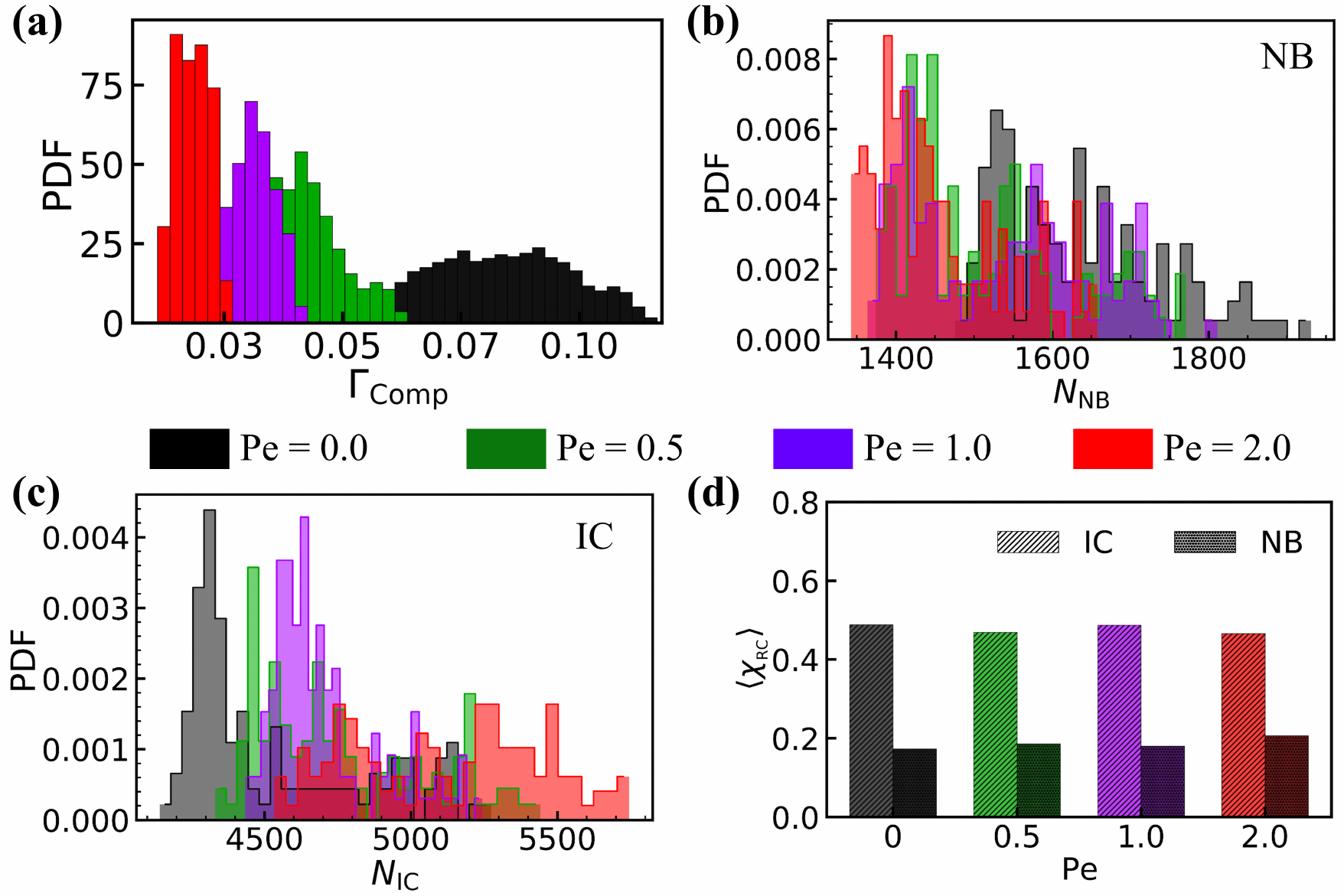}
    \caption{\textit{Increasing Pe percolates the network, enhancing interchain contact (IC) formation and retention while suppressing nonbonded intrachain (NB) contacts.} Probability distributions of (a) network compactness $\Gamma_\text{Comp}$, (b) NB, and (c) IC contacts, for $\rho_{_{M}} = 2.0$ with increasing $\mathrm{Pe}$. (d) Mean contact retention fraction $\left < \chi_{_\text{RC}} \right>$ as a function of $\text{Pe}$.}
    \label{fig:contact}
\end{figure}

\noindent To further probe the nature of the activity-driven reorganization of the condensate morphology, we construct a contact network in which each monomer is treated as a node, and edges are assigned between pairs of monomers that are either covalently bonded or within a cutoff distance $r_\text{cont}=1.0 \, \sigma_p$. Nonbonded contacts are further classified as intrachain (NB) or interchain (IC), depending on whether the two interacting monomers belong to the same chain or different chains, respectively. We define the network compactness $\Gamma_\text{Comp}$ as the average closeness centrality of the network, $\Gamma_\text{Comp} = \langle \mathcal{C}^{^\text{net}}_i \rangle$, where $\mathcal{C}^{^\text{net}}_i = (N-1)/\sum_j d_{ij}$ and $d_{ij}$ is the shortest-path distance between nodes $i$ and $j$.  As  $\text{Pe}$ increases, the distribution of $\Gamma_\text{Comp}$ shifts systematically to lower values (Fig.~\ref{fig:contact}a), confirming that the condensate evolves from a compact globule to a spatially extended, branched network. Consistent with this, the nonbonded intrachain contact distribution, $N_\text{NB}$, shifts to lower values with increasing $\text{Pe}$ (Fig.~\ref{fig:contact}b), indicating that activity suppresses self-contacts within individual chains. In contrast, the interchain contact distribution $N_\text{IC}$ broadens and shifts to higher values (Fig.~\ref{fig:contact}c), reflecting enhanced interchain connectivity that underpins the percolated network structure.\\

\noindent To assess the stability of these contacts, we define the contact retention fraction $\chi_{\mathrm{RC}}(t) = |C(t) \cap C(t+\Delta t)|/|C(t)|$, measuring the fraction of contacts that persist between consecutive frames, and is averaged over time. The mean contact retention fraction $\langle \chi_{_\text{RC}} \rangle$ remains consistently higher for IC contacts than for NB contacts across all $\text{Pe}$ (Fig.~\ref{fig:contact}d), demonstrating that interchain contacts are more persistently maintained, forming a stable network backbone even as individual contacts dynamically turn over. Further, the mean local ABP volume fraction, $\langle \phi_m \rangle$, computed within a sphere of radius $r=3.0 \, \sigma_s$ around each node, increases monotonically with $\textrm{Pe}$ for both IC and NB nodes, with nearly identical values (Fig.~S4(a, b)), indicating that ABPs remain broadly distributed throughout the condensate without preferential localization near NB or IC nodes. The node degree distribution, $\chi_{\rm node}$, further shows an enhanced population of higher-degree nodes with increasing Pe, reflecting the formation of branching network structures (Fig.~S4c). Similar trends are observed with increasing bath density, $\rho_{_{M}}$, including enhanced interchain connectivity, more highly branched nodes (Fig.~S5), larger $\langle \phi_m \rangle$, and persistently higher $\langle \chi_{_{\rm RC}} \rangle$ for IC than NB contacts (Fig.~S6).\\
\begin{figure*}[ht]
    \centering
    \includegraphics[width=0.9\textwidth]{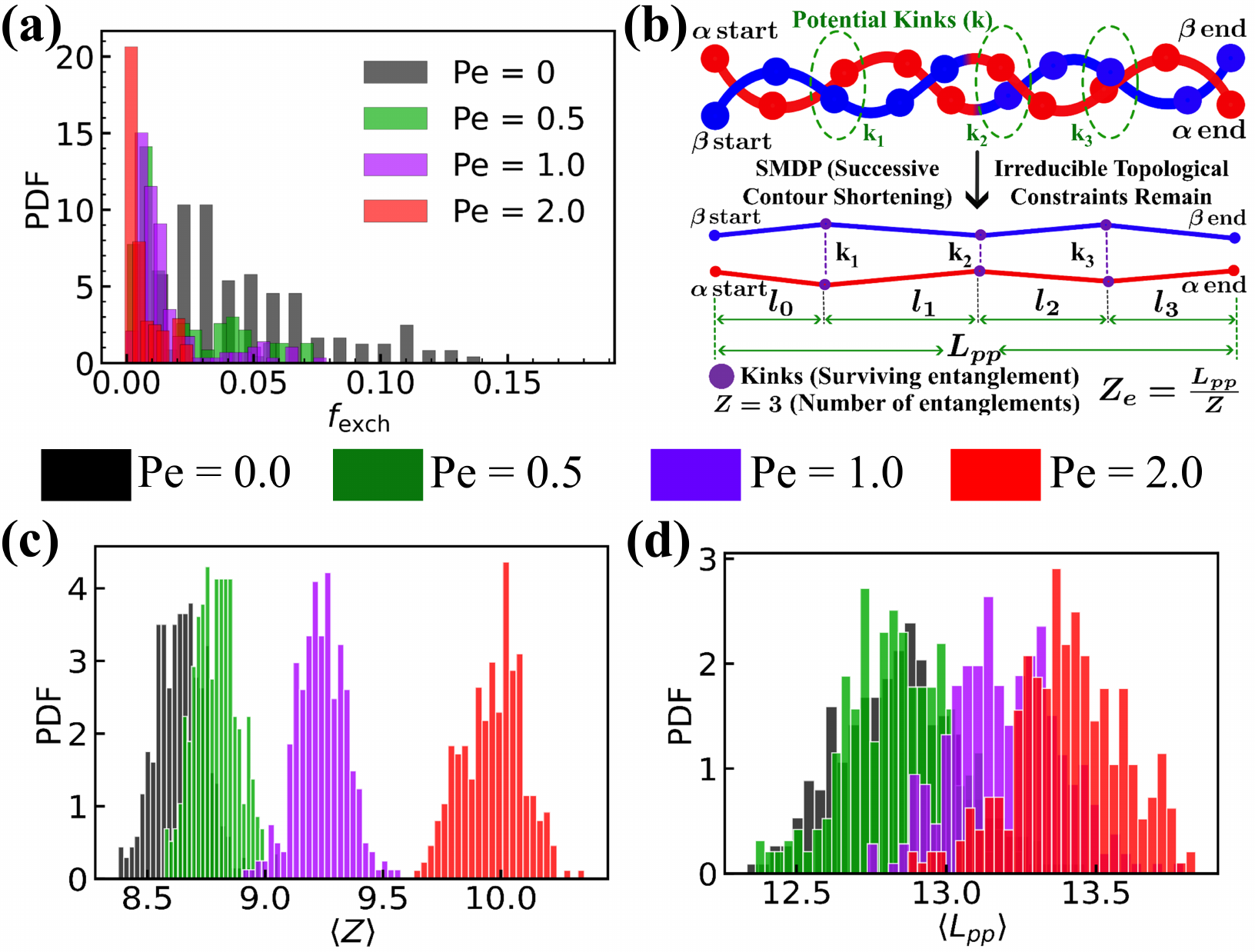}
    \caption{\textit{Dynamic chain exchange and topological robustness of the percolated network.} (a) Distribution of the exchange fraction $f_{\rm exch}$. (b) Schematic illustration of primitive path analysis using the SMDP algorithm for two entangled polymer chains ($\alpha$ and $\beta$). Distributions of (b) mean number of topological constraints per chain $\langle Z \rangle$, and (c) primitive path contour length $\langle L_{pp} \rangle$ for different $\text{Pe}$ ($\rho_{_{M}} = 2.0$).}
    \label{fig:exch_entang}
\end{figure*}

\noindent  Despite the formation of a structurally persistent percolated network, the condensate retains liquid-like character, as quantified by the exchange fraction $f_{\rm exch}$, which measures the fraction of polymer chains exchanged between the largest connected component (LCC) and the dilute phase. Polymer chains continuously exchange between the LCC and the dilute phase across all Pe (Fig.~\ref{fig:exch_entang}a), demonstrating that the network remains dynamically rearranging rather than kinetically arrested. The distribution of $f_{\rm exch}$ narrows and shifts toward lower values with increasing Pe (Fig.~\ref{fig:exch_entang}a), consistent with growth of the LCC (Fig.~S3a), which reduces the dilute-phase population available for exchange rather than suppressing exchange dynamics itself. Similar $f_{\rm exch}$ behavior is observed with increasing $\rho_{_{M}}$ (Fig.~S7a), confirming that the liquid-like character holds across bath densities. At the topological level, the robustness of the percolated network is further corroborated by primitive path entanglement analysis (Fig.~\ref{fig:exch_entang}b; see SM for details). The distributions of the mean number of constraints per chain, $\langle Z \rangle$, and the primitive path contour length, $\langle L_{pp} \rangle$, shift systematically to higher values with increasing Pe (Fig.~\ref{fig:exch_entang}(c, d)), indicating enhanced interchain winding and progressively more extended chain conformations within the percolated network. Correspondingly, the entanglement length, $\langle Z_e \rangle$, shifts to lower values with increasing Pe (Fig.~S7b), consistent with an increasing density of topological constraints. The increase in $\langle L_{pp} \rangle$ with Pe reflects the transition from locally folded chains in compact condensates to network-spanning bridges, where neighboring chains wind around bridging chains at topological junctions, stabilizing the branched structure (Fig.~S7c).\\
\begin{figure}[h!]
    \centering
    \includegraphics[width=0.48\textwidth]{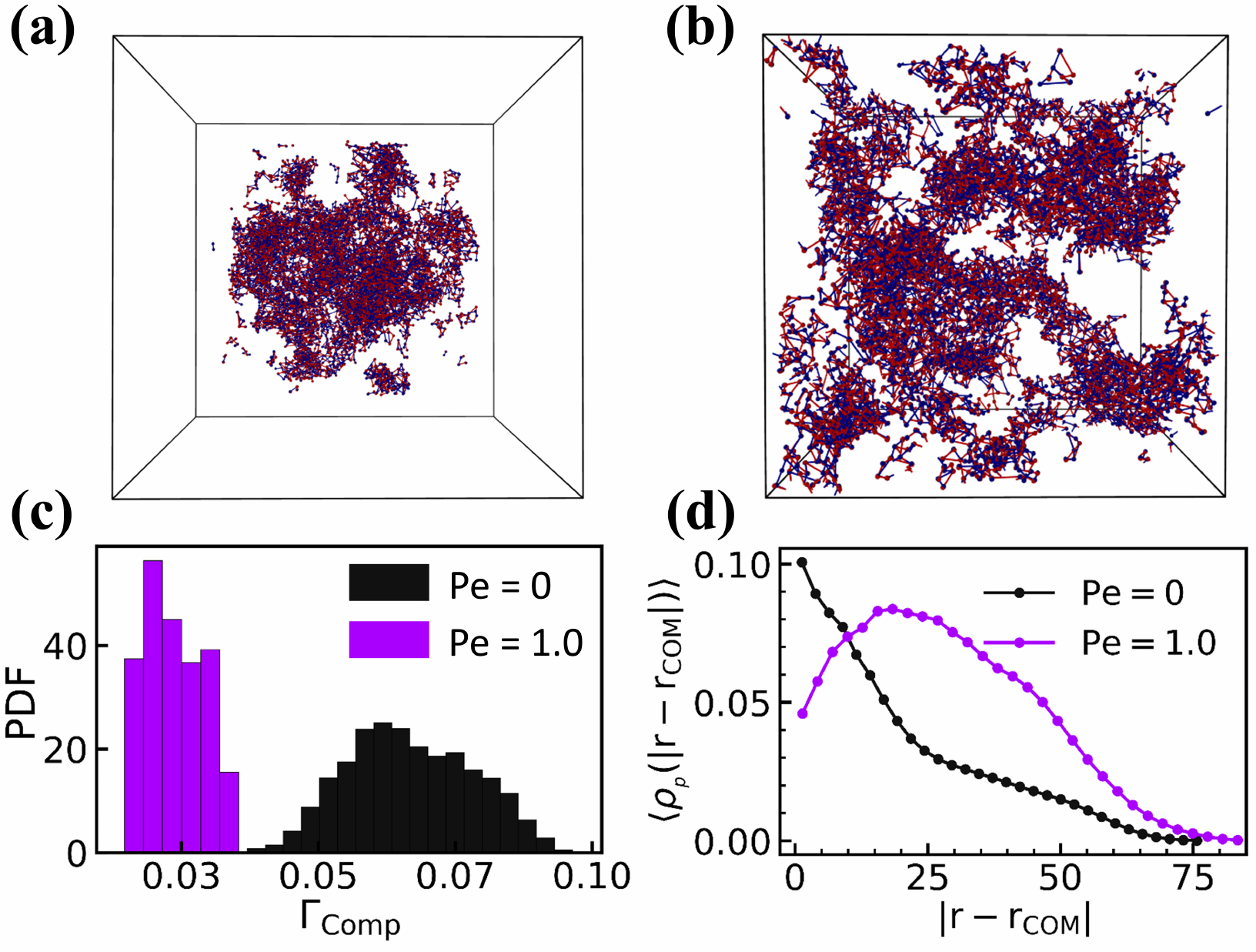}
    \caption{\textit{Activity-driven structural transition from a compact condensate to a spatially extended network.} Representative snapshots of the system for (a) $\text{Pe}=0$ and (b) $\text{Pe}=1.0$, showing a compact dense condensate and a spatially extended network structure, respectively. (c) Distribution of $\Gamma_\text{Comp}$ and (d) radial polymer density profile $\langle \rho_{_p}(|\text{r} - \text{r}_{\text{COM}}|) \rangle$ as a function of distance from the center of mass of the system ($\rho_{_{M}} = 2.0$).}
    \label{fig:stab-dep}
\end{figure}

\noindent The emergent percolated network raises a key question of whether its stability is maintained by the ABPs or is imprinted into the polymer contact topology itself. We test this via a controlled depletion of ABPs as a function of time after the network has formed (Fig.~S8). Specifically, ABPs are removed randomly in discrete steps during the depletion simulation, progressively reducing the active driving while leaving the polymer configuration unchanged, such that any subsequent structural changes arise solely from the loss of active fluctuations and polymer--ABP interactions. At each equally spaced interval, a fixed number $\Delta M = \lambda M_0$ of ABPs is deleted, where $M_0$ is the initial number of ABPs and $\lambda$ is the depletion rate, followed by a relaxation period before the next deletion step (Fig.~S8). After complete depletion of ABPs, the system is further evolved for a duration equal to twice the characteristic coalescence time of the polymer-only system, ensuring that any residual structural evolution toward a compact droplet is fully resolved. This protocol effectively quenches the system from an active steady state toward a passive limit, allowing us to determine whether the emergent network structure of the condensate is sustained by ongoing activity or remains stable in its absence.\\

\noindent The resulting state of the polymer condensate after completion of the ABP removal at a fixed rate $\lambda = 0.1$ is shown in Fig.~\ref{fig:stab-dep}. In the passive case ($\mathrm{Pe}=0$), depletion induces further compaction of the condensate, yielding a denser droplet than the steady-state configuration prior to depletion (Fig.~\ref{fig:stab-dep}a). In contrast, for an active bath ($\mathrm{Pe}=1.0$), the condensate retains a spatially extended, branched network morphology despite the removal of ABPs (Fig.~\ref{fig:stab-dep}b), indicating that the network structure persists beyond the active driving. This is reflected in the compactness distribution $\Gamma_{\text{Comp}}$, which is higher for $\mathrm{Pe}=0$ and shifted to lower values for $\mathrm{Pe}=1.0$ (Fig.~\ref{fig:stab-dep}c). This is further confirmed by the radial density profile for $\mathrm{Pe}=0$ is sharply peaked at the center and decays rapidly with distance, whereas for $\mathrm{Pe}=1.0$ it exhibits a broad peak at intermediate distances and decays more slowly, indicating redistribution of polymer density away from the center of mass and the formation of an extended architecture (Fig.~\ref{fig:stab-dep}d). Together, these results demonstrate that the network topology, once established by activity, remains stable even in the absence of ABPs, which differs from the findings in Ref.~\cite{gulati2026}.\\
\begin{figure}[h!]
    \centering
    \includegraphics[width=0.49\textwidth]{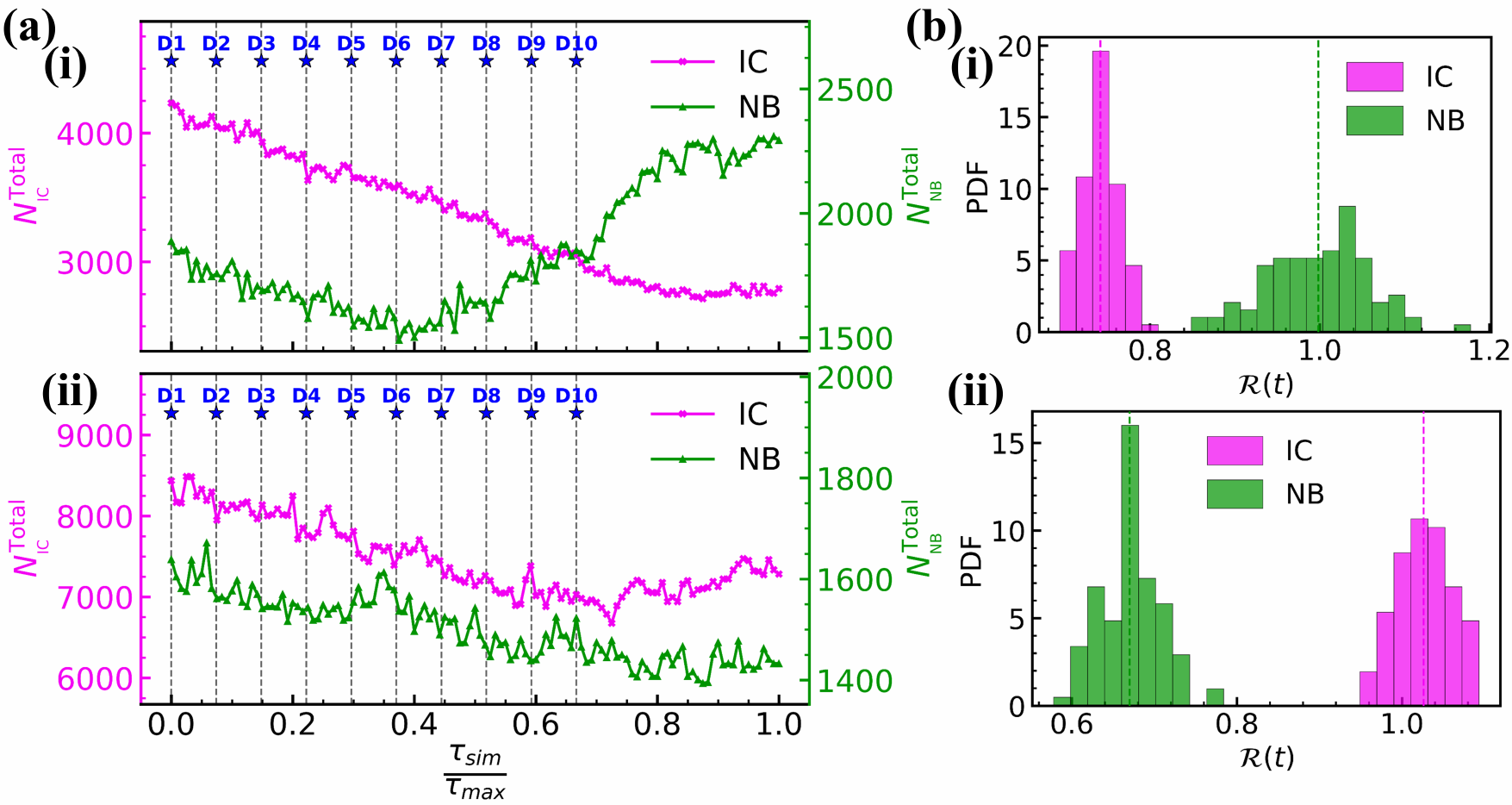}
    \caption{\textit{Activity drives a self-renewing condensate network that persists after depletion.} (a) Time evolution of IC and NB contacts and (b) renewal rate distributions $\mathcal{R}(t)$ for (i) $\mathrm{Pe}=0$ and (ii) $\mathrm{Pe}=1.0$. Dashed vertical lines mark successive ABP depletion ($\rho_{_{M}} = 2.0$, $\lambda = 0.1$) events D1-D10 in (a) and the mean $\mathcal{R}(t)$ for each contact type in (b).}
    \label{fig:contact_dynam}
\end{figure}

\noindent To elucidate the microscopic origin of this branched network stability, we examine the time evolution and renewal of IC and NB contacts during depletion (Fig.~\ref{fig:contact_dynam}). For $\mathrm{Pe}=0$, IC contacts decrease monotonically with each depletion event while NB contacts increase (Fig.~\ref{fig:contact_dynam}a-i), reflecting gradual compaction of the condensate into a compact droplet-like as ABPs are removed. In the active bath case, for $\mathrm{Pe}=1.0$, IC contacts remain largely stable, exhibiting a slight initial decrease followed by a subsequent increase after depletion, while NB contacts decrease (Fig.~\ref{fig:contact_dynam}a-ii), indicating preservation of the percolated network topology even as activity is withdrawn. The contact renewal rate $\mathcal{R}(t)$ further corroborates this picture: for $\mathrm{Pe}=0$, IC contacts have $\mathcal{R}(t)<1$, indicating net loss of interchain connectivity, whereas for $\mathrm{Pe}=1.0$, IC contacts maintain $\mathcal{R}(t) \approx 1$ while NB contacts are suppressed (Fig.~\ref{fig:contact_dynam}b), demonstrating that the condensate network is self-renewing and continuously replaces lost contacts to preserve its topology. These findings reveal that activity from the bath not only generates the percolated network but also imprints a dynamically stable, self-renewing interchain backbone that persists even after active driving is removed. Primitive path visualization confirms the persistence of topological winding points throughout the network bridges across all stages of depletion (Fig.~S9). Similar IC and NB contact dynamics are observed for faster depletion ($\lambda=0.4$), indicating that the percolated network remains stable against rapid withdrawal of activity (Fig.~S10).\\ 

\noindent We next examine the connectivity pattern underlying this self-maintained structure using the betweenness centrality, defined as $\mathcal{B}i = \frac{2}{(N_d - 1)(N_d - 2)}\sum\limits_{\substack{s < t \\ s,t \neq i}} \frac{\sigma_{st}(i)}{\sigma_{st}}$, where $\sigma_{st}$ is the total number of shortest paths between nodes $s$ and $t$, $N_d$ is the number of nodes in the connected component, and $\sigma_{st}(i)$ is the number of those paths that pass through node $i$. Thus, it quantifies how strongly a node $i$ mediates connectivity, with higher values indicating that a larger fraction of shortest paths pass through that node and are therefore less distributed across the topology. In the contact network of polymer condensates, nodes with degree $>2$ are classified as branch nodes, while those with degree 2 are linear nodes. Fig.~S11 displays the time evolution of the mean betweenness centrality $\langle \mathcal{B}_i \rangle$ for these two classes. Linear nodes maintain lower centrality than branch nodes throughout; in the passive case, they show modest fluctuations with occasional spikes near complete depletion, whereas in the active case, they remain smooth with weak temporal variation (Fig.~S11). This implies that they are predominantly associated with bonded polymer connections that are fixed, with only a small contribution from rearranging nonbonded contacts, leading to a limited role in structural reorganization. Branch nodes, by contrast, display markedly different behavior. For $\mathrm{Pe}=0$, branch nodes exhibit higher centrality with intermittent peaks that become more pronounced as the system approaches complete depletion ($N_{\mathrm{ABP}} \to 0$), highlighting that connectivity is concentrated in a small subset of nodes (Fig.~S11a). This facilitates structural rearrangement through these nodes, consistent with the observed collapse into a more compact droplet-like condensate relative to the initial steady state ($N_{\mathrm{ABP}} \neq 0$). At $\mathrm{Pe}=1.0$, $\langle \mathcal{B}_i \rangle$ decreases steadily over time, falling below the time average as $N_{\mathrm{ABP}}$ approaches zero (Fig.~S11b), indicating a more distributed connectivity with no dominant nodes. This redistribution suggests that activity promotes a more homogeneous and redundant architecture, where multiple pathways support global connectivity, corroborating the self-renewing contacts and enhanced stability of the percolated structure (Fig.~\ref{fig:stab-dep}b).\\

\noindent The emergence of a persistent percolated network cannot be explained solely by activity-induced swelling of individual polymer chains (Fig.~S12). Indeed, simulations with larger active Brownian particles (ABPs) exhibit an increase in the polymer radius of gyration~\cite{ching2025}, indicating more extended chain conformations, yet fail to produce a percolated network. These prior observations combined with our current observations here suggest that network formation requires a balance between polymer swelling and the generation of interchain topological constraints~\cite{chai2025dna}. When ABPs are too small, their excluded-volume interactions are insufficient to significantly open polymer conformations, and intrachain contacts continue to dominate. Conversely, when ABPs are too large, although they promote chain extension, they may sterically hinder the close interchain encounters and wrapping events needed to form stable bridging connections between polymers. We therefore propose that persistent network formation occurs within an intermediate regime where ABPs are sufficiently large to suppress local chain collapse, yet sufficiently small and mobile to facilitate interchain winding, bridge formation, and the accumulation of topological constraints. Moreover, changing the sticker-spacer motif can also influence the emergent structure. See Supplemental Materials for further details.\\

\noindent At the mesoscale, our system shares features with activated colloidal patchy-particle networks~\cite{jonas2025act}, where activity generates void-rich and dynamically connected structures. Similar to these systems, our polymer condensates maintain IC renewal rates close to $1$ ($\mathcal{R}(t) \approx 1$) while NB contacts are suppressed, reflecting a self-sustaining balance between contact formation and loss. The robustness of the percolated topology likely arises from multivalent sticker-spacer interactions that generate redundant connectivity pathways, a feature also central to PopZ condensates, where filamentous ultrastructure maintains condensate integrity and function~\cite{scholl2026}. Further, recent experiments on reconstituted multivalent protein condensates revealed that a percolating scaffold of non-mobile proteins sustains a dynamically exchanging mobile fraction~\cite{liao2025emergent}, similar to the coexistence of chain exchange dynamics and topological entanglement reported here. \\

\noindent Finally, activity-induced topological imprinting may provide a mechanism for storing information about past cellular states while preserving the adaptability and exchange characteristic of fluid condensates. Or, such a dynamically-maintained scaffold could, in biological contexts, help regulate access to binding sites or maintain spatial heterogeneity. Moreover, as the P\'{e}clet number increases even more so, we anticipate a transition to an arrested state, presumably similar to the earlier identified rigidity transition to establish a more standard pathway for memory. All in all, activity is capable of sculpting condensate architecture and function in nontrivial ways, particularly when considering condensates not just in active baths but also spatially confined by the cytoskeleton or other fiber-like environments~\cite{Style2018}. Indeed, droplets composed of network architectures may interact with environment networks in sophisticated ways. \\

\noindent \textit{Acknowledgments-} This research was supported in part through computational resources provided by Syracuse University. This research is funded in part by the Alfred P. Sloan Foundation under grant G-2024-22546 to JLR and JMS.

%\bibliography{Percolated_Condensates_Polymer}

%Supplemental Material
\clearpage
\onecolumngrid

\makeatletter
\long\def\@makecaption#1#2{%
  \vskip\abovecaptionskip
  \sbox\@tempboxa{#1: #2}%
  \ifdim \wd\@tempboxa >\hsize
    {\setlength{\rightskip}{0pt}%
     \setlength{\leftskip}{0pt}%
     \parfillskip=0pt plus 1fil\relax
     \noindent #1: #2\par}%
  \else
    \global\@minipagefalse
    \hb@xt@\hsize{\hfil\box\@tempboxa\hfil}%
  \fi
  \vskip\belowcaptionskip}
\makeatother

\section{\Large Supplemental Material: Active Particles Imprint Persistent Percolating Networks in Polymer Condensates}
\vspace{2em}

\renewcommand{\thefigure}{S\arabic{figure}}
\renewcommand{\figurename}{Fig.}
\setcounter{figure}{0}

\renewcommand{\thetable}{S\arabic{table}}
\setcounter{table}{0}
\captionsetup[table]{labelfont={bf},name={Table}}

\renewcommand\thesection{\arabic{section}}
\renewcommand\thesubsection{\thesection.\arabic{subsection}}

\section*{S\lowercase{imulation} D\lowercase{etails}}

\noindent We employ a sticker-spacer framework~\cite{choi2020physical, zhang2024exchange}, where each chain of sequence $(A_n B_n)_m$ consists of A and B type monomers of radius $\sigma_p$ serving as stickers, linked by finitely extensible nonlinear elastic (FENE) springs as implicit spacers~\cite{kremer1990dynamics}. The parameters $n$ and $m$ are varied at fixed total chain length ($N_p = 12$) to isolate the effect of sticker distribution along the chain. The active bath is modeled as $N_s$ spherical active Brownian particles (ABP) with radius $r_s = 1.5$ nm. The implicit spacers are modeled using the following spring potential:
\begin{equation}
V_{\textrm{b}}(r_{ij}) = - \frac{1}{2} k_{b} R_0^2 ln \left[ 1 - \left( \frac{r_{ij}}{R_0} \right)^2\right], \hspace{5mm}  r_{ij} < R_0
\label{eq:bond}
\end{equation}
where $r_{ij}$ is the distance between the $i^{th}$ and $j^{th}$ stickers on the polymer chain, $k_b$ is the spring constant, and $R_0^2$ is the maximum extension of the bond. The interactions between monomers are determined by the sticker types. Sticker types A-B interact via a soft attractive potential that models specific, one-to-one binding with affinity strength, $\epsilon_{AB}$. The soft attractive potential is defined as:
\begin{equation}
V_{\text{A}}(r_{ij}) = - \epsilon_{AB}  \left( 1 + cos \frac{\pi r_{ij}}{r_c}\right),  \hspace{5mm}  r_{ij} < r_c
\label{eq:soft}
\end{equation} 
where $\epsilon_{AB}$ is the strength of the attractive interaction or the binding affinity and $r_c$ is the cutoff distance. All remaining pairs, including same-type monomers (A-A and B-B), monomer-ABP, and ABP-ABP interactions, are governed by a purely repulsive Weeks--Chandler--Andersen (WCA) potential~\cite{weeks1971role}, preventing nonspecific aggregation and polymer collapse. The WCA potential is given by:
\begin{equation}
V_{\text{R}}(r_{ij}) =
\begin{cases}
4\epsilon_{AA} \left[ \left( \dfrac{\sigma_{ij}}{r_{ij}} \right)^{12} - \left( \dfrac{\sigma_{ij}}{r_{ij}} \right)^6 \right] + \epsilon_{AA}, & r_{ij} \leq 2^{1/6}\sigma_{ij} \\
0, & r_{ij} > 2^{1/6}\sigma_{ij}
\end{cases}
\label{eq:wca}
\end{equation}
where $r_{ij}$ is the distance between the $i^{\text{th}}$ and $j^{\text{th}}$ particles (polymers or the bath particles), $\sigma_{ij} = \frac{\sigma_i + \sigma_j}{2}$ is the interaction range with $\sigma_{i(j)}$ denoting the diameter of the $i^{\text{th}}$ ($j^{\text{th}}$) particle, and $\epsilon_{AA}$ is the interaction strength. \\

\noindent All simulations are performed using LAMMPS~\cite{plimpton1995fast} software package. We consider $k_b = 0.15 \, k_B T/\mathrm{nm}^2$, $R_0 = 10 \, \mathrm{nm}$, $\epsilon_{AB} = 6\, k_B T$, and $r_c = 1\, \mathrm{nm}$, with particle diameters $\sigma_p = 2\, \mathrm{nm}$ for polymers and $\sigma_s = 3\, \mathrm{nm}$ for bath particles in all simulations (Table~\ref{tab:parameters_sim}). The system contains $N = 1000$ polymer chains in a cubic simulation box of edge length $107.72\, \mathrm{nm}$ with periodic boundary conditions, corresponding to a polymer volume fraction of $\phi_p =  0.04$. The bath particles are introduced into the system with a composition characterized by the ABP fraction, $\rho_{_{M}} = \frac{N_\mathrm{ABP}}{N_\mathrm{mon}}$, where $N_\mathrm{mon}$ is the total number of polymer monomers, with $\rho_{_{M}}$ varied as \{0.5, 1.0, 1.5, 2.0, 2.5\} ensuring the total packing fraction of polymers and ABPs remains below $0.35$ across all simulations. The dynamics evolve according to the overdamped Langevin equation:
\begin{equation}
\gamma \, \dot{\bm{r}}_i = - \nabla_i V(\bm{r}_{1},\bm{r}_{2},..., \bm{r}_{N}) + \bm{f}_{i}(t) + F_\mathrm{act} \bm{\hat{n}}_i  \label{eq:langevineq}
\end{equation}
where $\bm{r}_i$ denotes the position of the $i$th particle, $\gamma$ is the friction coefficient, and $V(\bm{r}_1, \ldots, \bm{r}_N)$ accounts for all bonded and non-bonded interaction potentials. The stochastic force term $\bm{f}_{i}(t)$ represents Gaussian white noise satisfying the fluctuation-dissipation theorem, and the term $F_\mathrm{act} \bm{\hat{n}}_i$ is the self-propulsion force acting exclusively on bath particles, where $F_\mathrm{act}$ is the magnitude of the activity, and $\bm{\hat{n}}_i $ denotes the unit orientation vector of particle $i$. The activity is applied exclusively on bath particles, with each ABP experiencing a constant force of magnitude $F_\mathrm{act}$ directed along its instantaneous orientation $\bm{\hat{n}}_i $, while polymer monomers remain passive throughout. The orientation vector $\bm{\hat{n}}_i $ evolves via rotational diffusion, implemented through stochastic quaternion updates, $\frac{d\bm{\hat{n}}_i}{dt} = \bm{\xi}_i(t) \times \bm{\hat{n}}_i,$ where $\bm{\xi}_i(t)$ is a Gaussian white-noise vector with $\langle \bm{\xi}_i(t) \rangle = 0$, and correlations $\langle \xi_{\alpha}(t)\xi_{\beta}(t') \rangle = 2D_r \delta_{\alpha\beta}\delta(t - t')$, where $D_r$ is the rotational diffusion coefficient setting the persistence time of active motion. The strength of activity is expressed in terms of the dimensionless P\'{e}clet number:
\begin{equation}
Pe = \frac{F_\mathrm{act} \sigma_s}{k_B T}
\end{equation}
where $\sigma_s = 3 \, nm$is the diameter of the bath particles, with $\mathrm{Pe} \in \{0, 0.5, 1.0, 2.0\}$ spanning passive to strongly active conditions. All simulations are performed using LAMMPS~\cite{plimpton1995fast} software package at $T = 300\, \mathrm{K}$, with a damping timescale $m/\gamma = 10\, \mathrm{ns}$ and integration timestep, $dt = 5 \times 10^{-4}\, \mathrm{ns}$. Full details of parameter choices are provided in the Table~\ref{tab:parameters_sim}.
\begin{table}[h]
\centering
\caption{Parameters used in the simulations}
\label{tab:parameters_sim}
  \begin{tabular*}{0.9\textwidth}{@{\extracolsep{\fill}}lll}
    \hline
    Parameters & Numerical Value \\
    \hline
    Number of polymer monomers per chain ($N_p$) & 12  \\
    Number of polymers ($N$) & 1000  \\
    Fraction bath particle ($\rho_{_{M}}$) & 0.5, 1.0, 1.5, 2.0, 2.5  \\
    Polymer monomer diameter ($\sigma_p$) & 2 nm  \\
    Diameter of a bath particle ($\sigma_s$) & 3 nm  \\
    Simulation box dimension ($V_\mathrm{box}$) & 107.72 $\times$ 107.72 $\times$ 107.72 $\mathrm{nm}^3$\\
    Total polymer volume fraction ($\phi_{p}$) & 0.04 \\
    Pecl\`et  number ($Pe$) & 0, 0.5, 1.0, 2.0 \\
    Temperature ($T$) & 300 K \\
    Spring constant ($k_b$) & 0.15 $k_BT/\mathrm{nm}^2$ \\
    Maximum extent of the bond ($R_0$ ) & 10 nm \\
    Binding affinity ($\epsilon_{AB}$) & 6 $k_BT$ \\
    Cutoff for soft attraction  ($r_c$) & 1 nm \\
    WCA interaction strength ($\epsilon_{AA}$) & 1 $k_BT$ \\ 
    Damping timescale ($\frac{m}{\gamma}$) & 10 ns \\
    Total simulation time ($\tau_{\mathrm{sim}}$) & 40000 $\tau$ \\
    Simulation timestep ($d\tau$) & $ 5 \times 10^{-4}$ ns \\
    \hline
  \end{tabular*}
\end{table}

\noindent The simulation is initialized by placing polymers and bath particles randomly in the box, followed by a short relaxation run of $10^7$ steps with only excluded volume interactions to randomize the configuration, before gradually ramping the attractive interaction between A and B stickers from 0 to $\epsilon_{AB}$ over $15 \times 10^6$ integration steps. For $\mathrm{Pe}\neq 0$, activity is subsequently applied to the bath particles, and the system is further relaxed for $5 \times 10^6$ steps to reach a steady state, while for $\mathrm{Pe} = 0$,  the simulation is continued for the same duration to maintain equal total simulation time across all conditions. The system is further evolved for an additional $25 \times 10^6$  steps in the steady state, during which particle configurations are recorded every $25\times 10^3$ steps for analysis. We perform five independent simulation replicates for each combination of  $\mathrm{Pe}$, $\epsilon_{AB}$, and $\rho_{_{M}}$, and all reported quantities are averaged over these replicates.\\

\begin{figure}[h!]
    \centering
    \includegraphics[width=0.65\textwidth]{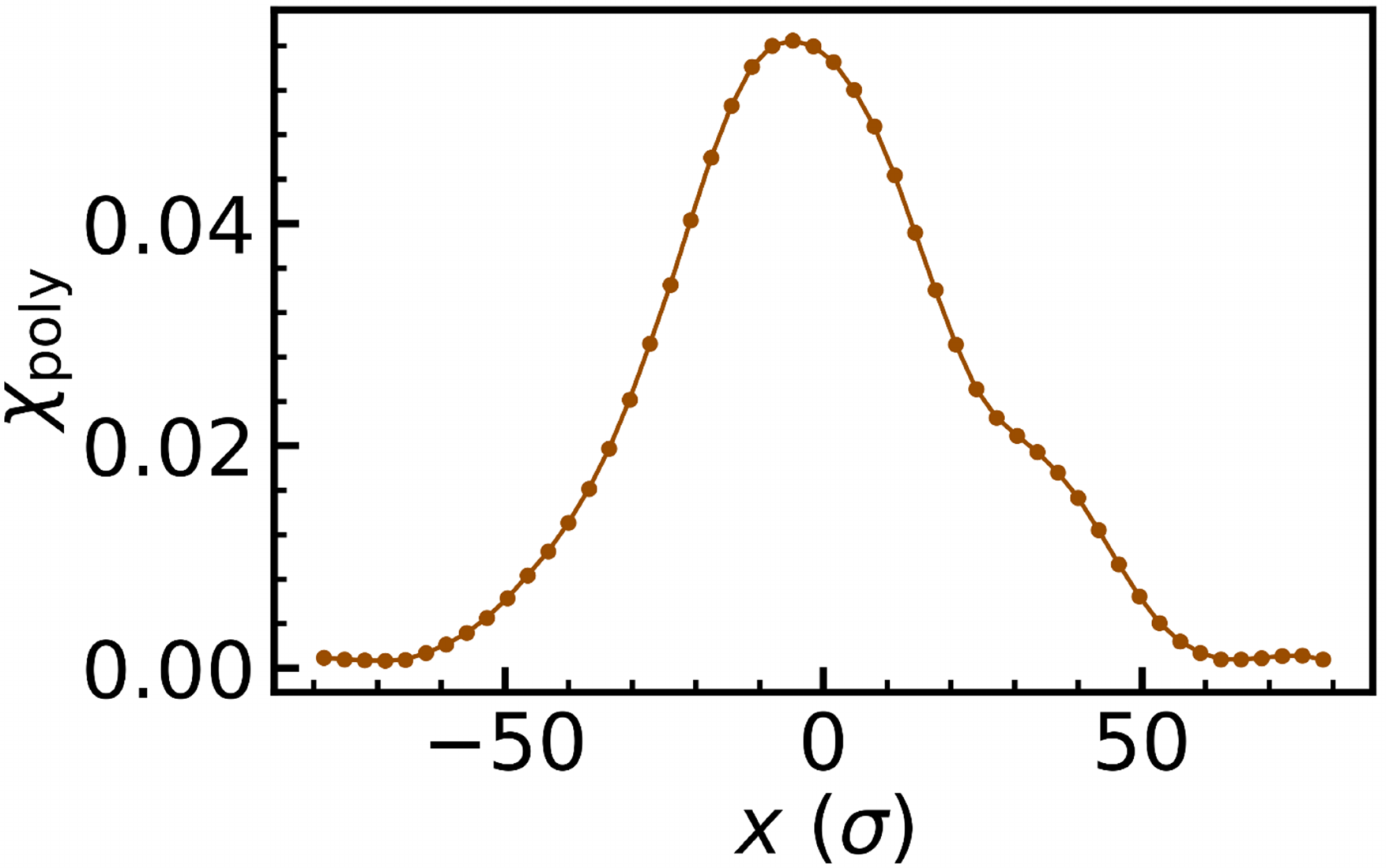}
    \caption{\small \textit{Formation of a compact droplet through coarsening in the polymer-only case.} Monomer concentration profile of polymers along the x-axis of the simulation box.}
    \label{fig:conprofile}
\end{figure}

\noindent For the polymer-only system, the spatial polymer volume fraction, $\chi_{\rm poly}$, is obtained by binning monomer positions along the $x$ direction into 50 equally spaced bins spanning the simulation box as shown in Fig.~\ref{fig:conprofile}. The monomer count in each bin is normalized by the total number of monomers and averaged over simulation frames after recentering the system on its center of mass. The resulting unimodal profile reflects polymer enrichment within the dense condensate near $x = 0 \,\sigma$, with $\chi_{\rm poly}$ decaying to near zero in the dilute phase at large $|x|$ (Fig.~\ref{fig:conprofile}).\\

\noindent\textbf{System-spanning analysis and fraction of chains in the largest connected component (LCC):} To quantify the spatial extent and connectivity of the percolated polymer network, we perform two complementary analyses a system-spanning (percolation) calculation and a chain-level largest connected component (LCC) analysis.\\

\noindent A contact network is constructed in which each monomer is treated as a node, and edges are assigned between pairs of monomers within a cutoff distance $r_{\rm cont} = 1.0\,\sigma_p$. Bead coordinates are unwrapped prior to graph construction to ensure chain continuity across periodic boundaries. LCC is identified as the connected subgraph with the maximum number of nodes and edges. The fraction of polymer chains participating in LCC,, is computed by assigning each chain $c$ to the LCC if at least one of its monomers belongs to LCC (Fig.~\ref{fig:span}a) as:
\begin{equation}
    f^{\rm LCC}_{\rm chain}(t) = \frac{n_{\rm LCC}(t)}{N_{\rm chains}}
    \label{eq:frac_lcc}
\end{equation}
\noindent where $n_{\rm LCC}(t)$ is the number of chains that have at least one monomer in LCC at snapshot $t$ and $N_{\rm chains}$ is the total number of polymer chains. The steady-state ensemble average $\langle f^{\rm LCC}_{\rm chain} \rangle = \frac{1}{N_{\rm snap}}\sum_t f^{\rm LCC}_{\rm chain}(t)$ is shown in Fig.~\ref{fig:span}a.\\

\noindent The spatial extent of LCC, $S_\alpha$, is quantified by the span in each direction $\alpha \in \{x, y, z\}$, defined as the range of unwrapped bead coordinates of the LCC normalized by the box length $L$:
\begin{equation}
    S_\alpha = \frac{r_\alpha^{\rm max} - r_\alpha^{\rm min}}{L}
    \label{eq:span}
\end{equation}

\noindent where $r_\alpha^{\rm max}$ and $r_\alpha^{\rm min}$ are the maximum and minimum unwrapped coordinates of beads in LCC along direction $\alpha$. For the cubic simulation box, the isotropic span per snapshot is $\mathcal{S} = (S_x + S_y + S_z)/3$, and the reported bar chart value is the steady-state ensemble average $\langle \mathcal{S} \rangle$ computed over all analyzed snapshots (Fig.~\ref{fig:span}b). The network is classified as system-spanning (percolated) when the LCC fills more than $85\%$ of the simulation box simultaneously in all three directions, i.e., $S_\alpha > S_{\rm threshold} = 0.85$ for $\alpha \in \{x, y, z\}$. Clusters extending across periodic boundaries yield $S_\alpha > 1$, correctly reflecting system-spanning connectivity.\\

\noindent\textbf{Entanglement analysis:} Entanglement analysis was performed using the Z1+ package~\cite{Z1+2023}, which implements the shortest multiple disconnected path (SMDP) algorithm. For each polymer chain $\alpha$ in a system of $N$ chains, the algorithm determines the primitive path (PP), defined as the shortest path connecting the fixed chain ends while preserving all topological constraints with neighboring chains. The primitive path is obtained through an iterative contour-shortening procedure in which interior vertices are successively simplified wherever possible without allowing chains to cross (intuitively, this can be viewed as interior beads being successively removed whenever their removal does not require a chain to cross a neighboring chain). Segments that can be shortened without altering the topology are eliminated, whereas regions constrained by neighboring chains cannot be further reduced. These irreducible topological constraints appear as kink (entanglement) points on the primitive path (Fig.~X). The number of interior kink points on the primitive path of chain $\alpha$ defines its entanglement number, $Z_\alpha$, and the ensemble-averaged entanglement is given by:
\begin{equation}
\langle Z \rangle = \frac{1}{N}\sum_{\alpha=1}^{N} Z_\alpha
\end{equation}
The primitive path contour length $L_{pp}^\alpha$ is computed as the sum of straight-line distances between successive entanglement points $\mathbf{p}_{\alpha, k}$ from one chain end to the other, and the ensemble average over all $N$ polymer chains $\langle L_{pp}\rangle $ are given by:
\begin{equation}
L_{pp}^\alpha = \sum_{k=0}^{Z_\alpha} |\mathbf{p}_{\alpha, k+1} - \mathbf{p}_{\alpha, k}|, \qquad
\langle L_{pp}\rangle = \frac{1}{N}\sum_{\alpha=1}^{N} L_{pp}^\alpha
\end{equation}
The entanglement length $Z_e^\alpha$ is defined as the average contour length of the primitive path between successive entanglement points, and the ensemble average over all polymer chains $\langle Z_e \rangle$ is defined as:
\begin{equation}
Z_e^\alpha = \frac{L_{pp}^\alpha}{Z_\alpha}, \quad Z_\alpha > 0, \qquad
\langle Z_e \rangle = \frac{1}{N^\prime}\sum_{\alpha, Z_{\alpha}>0} Z_e^\alpha
\end{equation}
where $N^\prime = |{\alpha:Z_\alpha > 0}|$  is the number of entangled chains, and the sum excludes unentangled chains for which $Z_e^\alpha$ is undefined. All three quantities are computed by time-averaging over the steady-state trajectory frames. The distributions of $\langle Z \rangle$ and $\langle L_{pp} \rangle$ are reported in the main text (Fig.~3(b, c)). The entanglement length $\langle Z_e \rangle$ shifts to lower values with increasing Pe (Fig.~\ref{fig:entang}), consistent with the number of entanglement constraints per chain $\langle Z \rangle$ growing proportionally faster than the primitive path length $\langle L_{pp} \rangle$, indicating that entanglement points become more densely spaced along the chain backbone as activity drives the formation of the percolated network.\\

\noindent\textbf{Visualization of primitive paths and entanglement points:} To visualize the entanglement topology of the percolated network, we export the Z1+ primitive path output using the \texttt{Z1+export} utility, which generates output files containing both the original chain configurations and their primitive paths, and visualize them in OVITO~\cite{stukowski2010visualization}. The original polymer chains are rendered in gray to show the underlying chain connectivity, while the primitive path is rendered in a distinct color overlaid on the original chains. The primitive path terminal nodes, which coincide with the fixed chain ends, are shown in blue, and the interior kink points, where a neighboring chain winds around the bridging chain, preventing removal during SMDP reduction and thus marking a topological winding point, are highlighted in red. The spatial distribution of red kink points throughout the network bridges, rather than being confined within individual condensate nodes, provides direct visual evidence that the branches connecting condensate nodes are themselves topologically wound, reinforcing the structural stability of the percolated network (Fig.~\ref{fig:entang}b and Fig.~\ref{fig:entang_depletion}). \\

\begin{figure}[h!]
    \centering
    \includegraphics[width=0.985\textwidth]{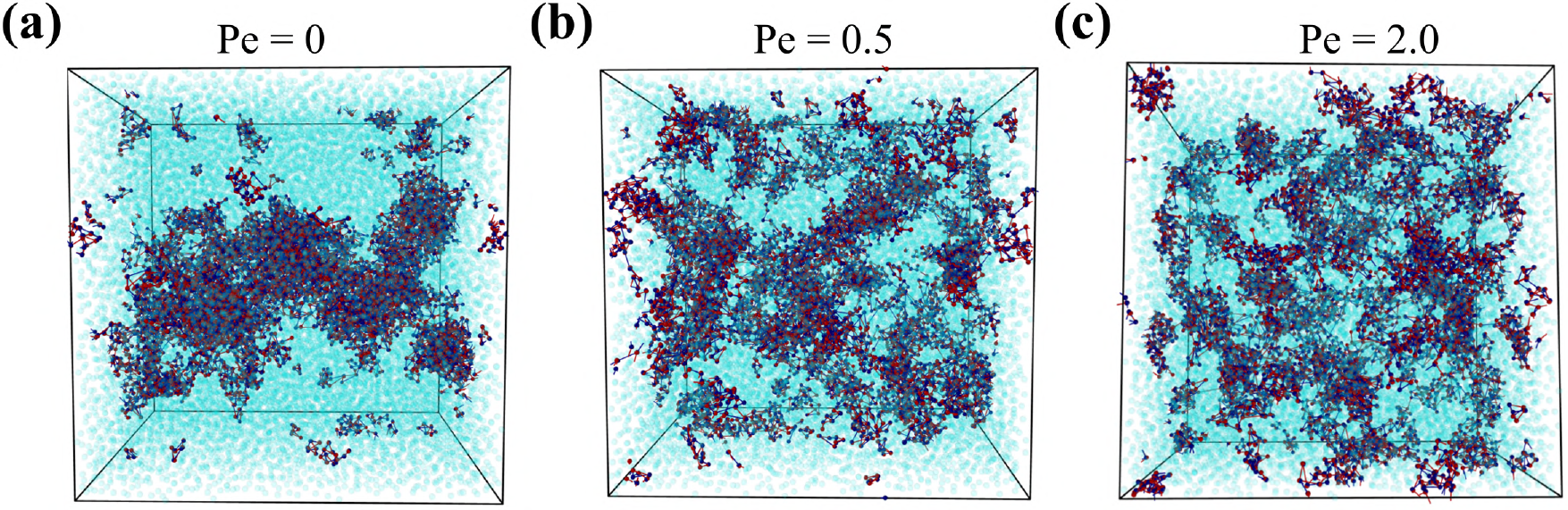}
    \caption{\small \textit{Activity-driven transition from compact condensate to percolated networks.} Representative simulation snapshots at (a) $\text{Pe}= 0$, (b) $\text{Pe}= 0.5$, and (c) $\text{Pe} = 2.0$ for $\rho_{_{M}} = 2.0$. Red and blue beads represent the two monomer types of the polymer chains. Cyan particles are the bath particles.}
    \label{fig:snapshot}
\end{figure}
\begin{figure}[h!]
    \centering
    \includegraphics[width=0.95\textwidth]{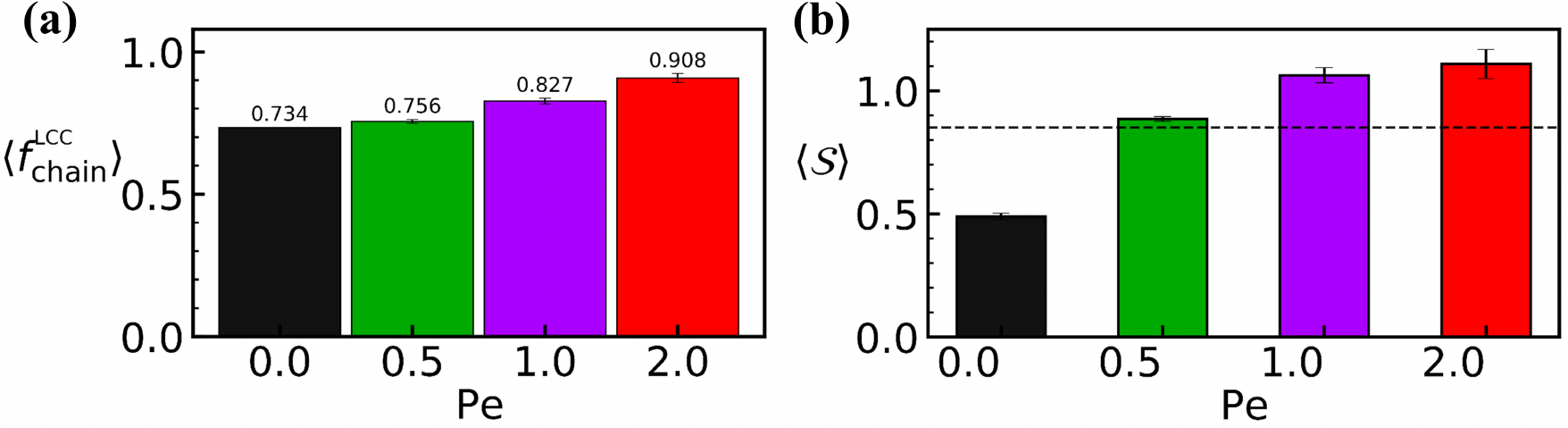}
    \caption{\small\textit{Activity-driven growth of the largest connected component (LCC) and system-spanning extent (percolation).} (a) Steady-state ensemble average of the fraction of polymer chains in the LCC, $\langle f_{\rm chain}^{\rm LCC} \rangle$ and (b) mean isotropic span $\langle \mathcal{S} \rangle$ of the LCC normalized by the box length, as a function of $\text{Pe}$ ($\rho_{_{M}} = 2.0$). The dashed line marks the percolation threshold $S_{\rm threshold} = 0.85$; values above this threshold indicate a system-spanning network.}
    \label{fig:span}
\end{figure}
\begin{figure}[h!]
    \centering
    \includegraphics[width=0.8\textwidth]{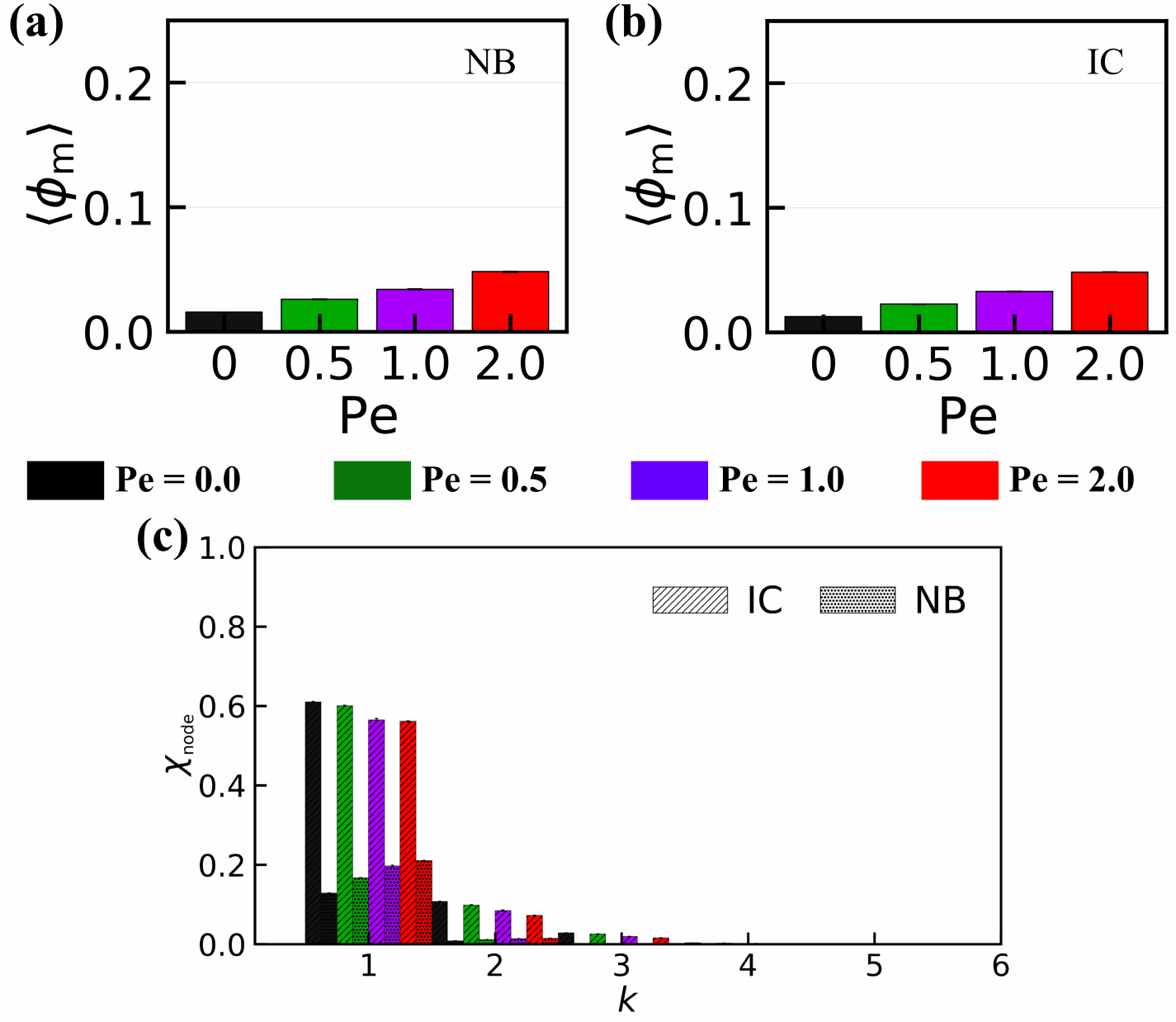}
    \caption{\small\textit{ABP localization near network nodes and branching statistics.} Mean local ABP volume fraction around (a) nonbonded intrachain (NB) and (b) interchain (IC) nodes as a function of $\text{Pe}$ for $\rho_{_{M}} = 2.0$. (c) Degree distribution of network nodes, $\chi_{\rm node}$, for different $\text{Pe}$ ($\rho_{_{M}} = 2.0$).}
    \label{fig:abpcon_pe}
\end{figure}
\begin{figure}[h!]
    \centering
    \includegraphics[width=0.7\textwidth]{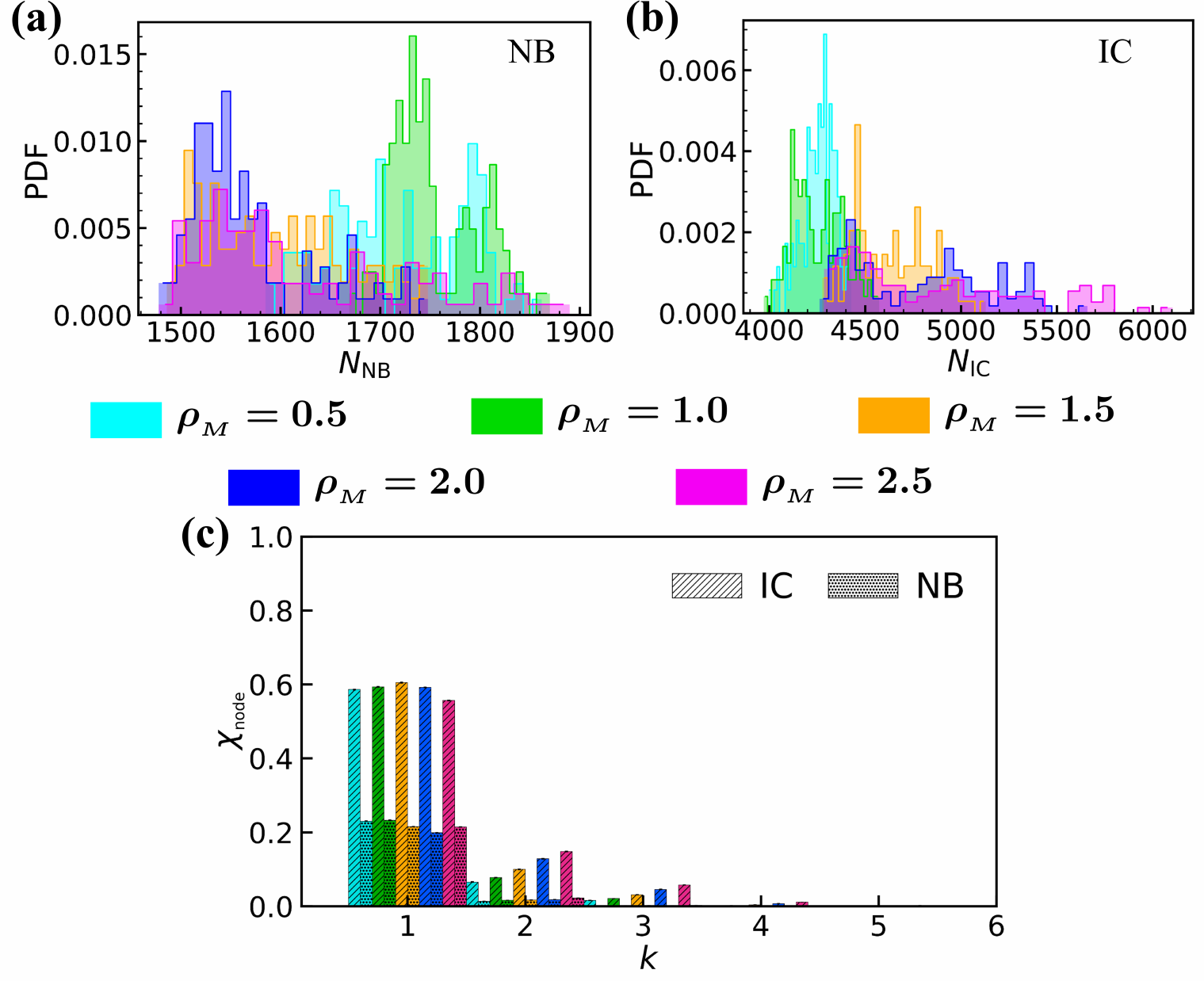}
    \caption{\small\textit{Nonbonded intrachain (NB) and interchain (IC) contacts for varying ABP fraction ($\rho_{_{M}}$).} Probability distributions of (a) NB and (b) IC contacts, for $\mathrm{Pe}=1.0$ with increasing $\rho_{_{M}}$. (c) Degree distribution of network nodes, $\chi_{\rm node}$, as a function of $\rho_{_{M}}$ for $\mathrm{Pe}=1.0$.}
    \label{fig:contact_rho}
\end{figure}
\begin{figure}[h!] 
    \centering
    \includegraphics[width=0.6\textwidth]{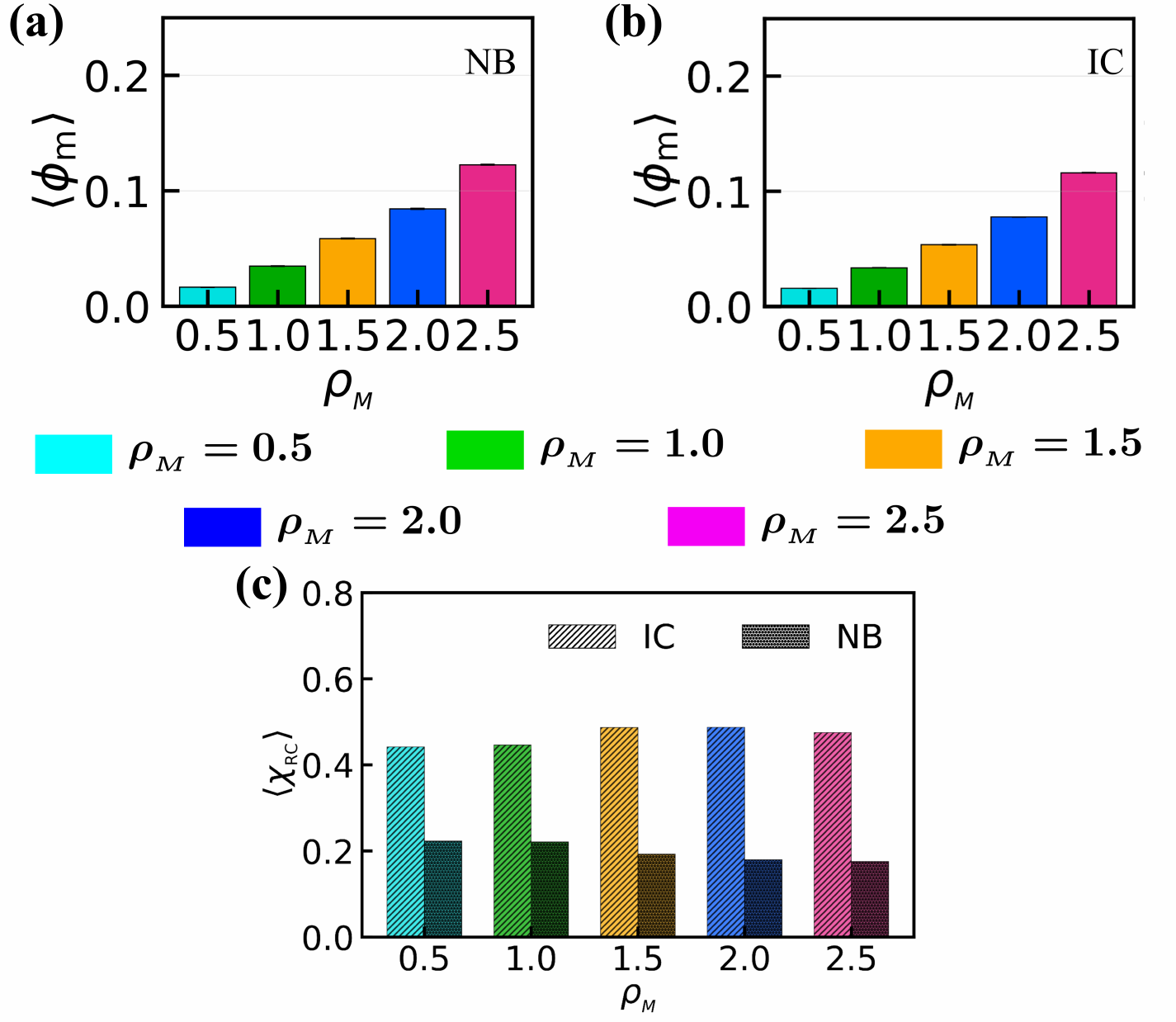}
    \caption{\small\textit{ABP localization near network nodes and contact retention.} Mean local ABP volume fraction around (a) nonbonded intrachain (NB) and (b) interchain (IC) nodes for $\mathrm{Pe}=1.0$ with increasing $\rho_{_{M}}$. (c) Mean contact retention fraction $\left < \chi_{_\text{RC}} \right>$ as a function of $\rho_{_{M}}$ for $\mathrm{Pe}=1.0$.}
    \label{fig:abpcon_rho}
\end{figure}
\begin{figure}[h!]
    \centering
    \includegraphics[width=0.975\textwidth]{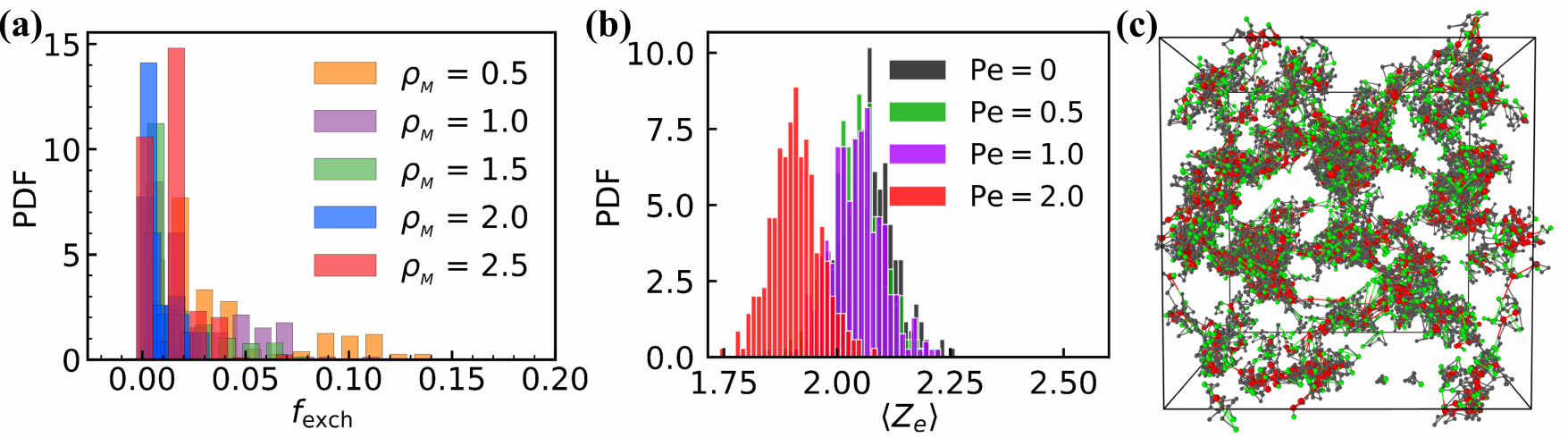}
    \caption{\small\textit{Dynamic exchange and entanglement topology of the percolated condensate network.} Probability distribution of the (a) chain exchange fraction $f_{\rm exch}$ and (b) mean entanglement length $\langle Z_e \rangle$ for $\rho_{_{M}} = 2.0$ at varying Pe. (c) Representative simulation snapshot for $\text{Pe} = 1.0$ ($\rho_{_{M}} =  2.0$) with primitive paths overlaid. Gray beads and bonds represent the original polymer chain configurations forming the percolated network. Green beads mark the primitive path terminal nodes, and red beads denote the entanglement (topological winding) points, locations where a neighboring chain winds around the bridging chain.}
    \label{fig:entang}
\end{figure}
\begin{figure}[h!]
    \centering
    \includegraphics[width=0.9\textwidth]{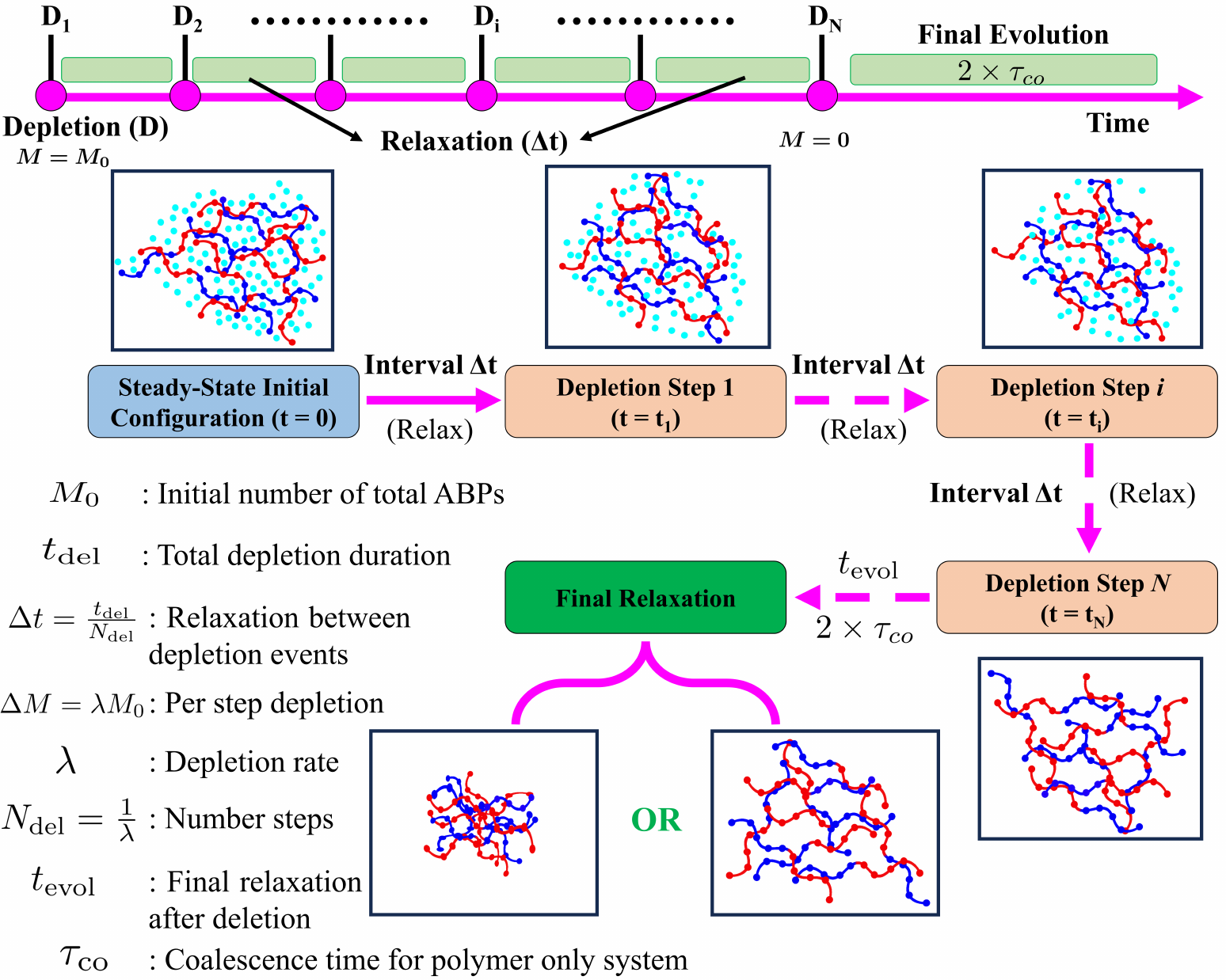}
    \caption{\small\textit{Schematic of the controlled stepwise ABP depletion protocol.} Starting from a steady-state configuration ($M = M_0$), a fixed number of ABPs, $\Delta M = \lambda M_0$, is removed at each of $N_{\mathrm{del}} = 1/\lambda$ discrete depletion events, separated by a relaxation interval $\Delta t = t_{\mathrm{del}}/N_{\mathrm{del}}$, until all ABPs are depleted ($M=0$). The system is then evolved for a final relaxation period $t_{\mathrm{evol}} = 2\tau_{\mathrm{co}}$, twice the coalescence time of the polymer-only system, yielding either a dense, compact droplet ($\mathrm{Pe} = 0$) or a persistent percolated network ($\mathrm{Pe} > 0$), depending on the activity.}
    \label{fig:dep-protocol}
\end{figure}
\begin{figure}[h!]
    \centering
    \includegraphics[width=0.975\textwidth]{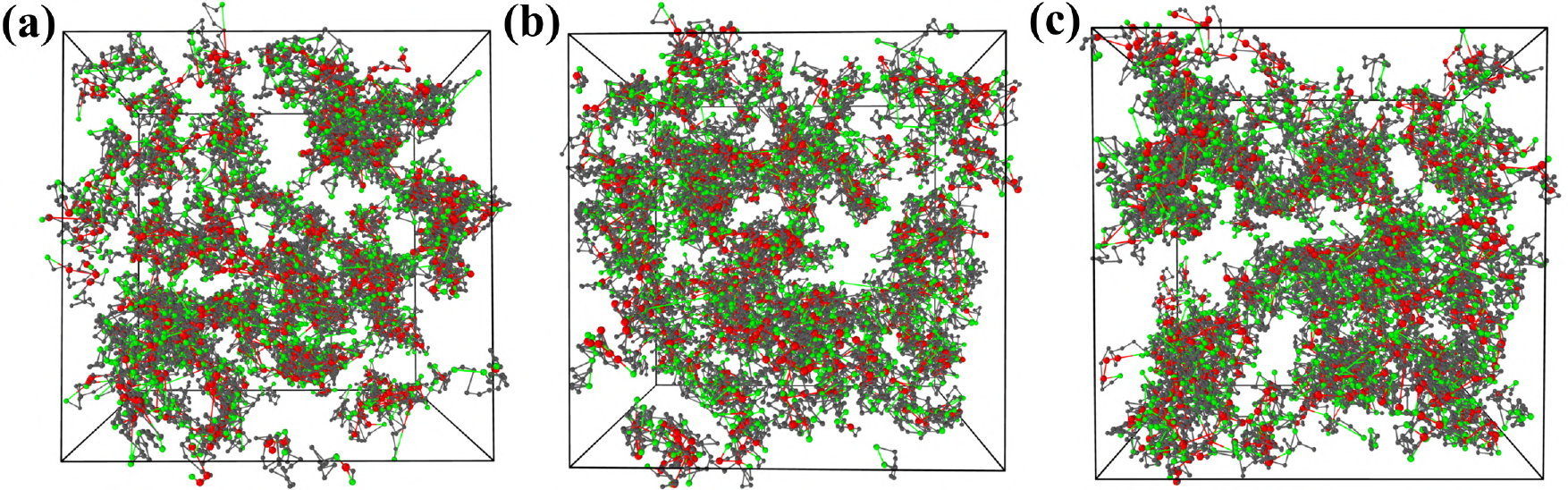}
    \caption{\small\textit{Primitive path visualization during ABP depletion.} Snapshots of the polymer condensate network with overlaid primitive paths at three stages of ABP depletion ($\lambda = 0.1$) for $\mathrm{Pe} = 1.0$ and $\rho_{_{M}} = 2.0$: (a) 50\% depletion, (b) immediately after complete depletion, and (c) after relaxation following complete depletion, corresponding to twice the coarsening time of the polymer-only system. Original polymer chains are rendered in gray, primitive path terminal nodes in green, and entanglement points marking topological winding locations in red. Across all three panels, the system-spanning percolated network topology is preserved without coarsening into a compact droplet, demonstrating that the interchain topological winding and enhanced connectivity driven by active fluctuations impart mechanical stability to the network that persists even after complete ABP depletion.}
    \label{fig:entang_depletion}
\end{figure}
\begin{figure}[h!]
    \centering
    \includegraphics[width=0.75\textwidth]{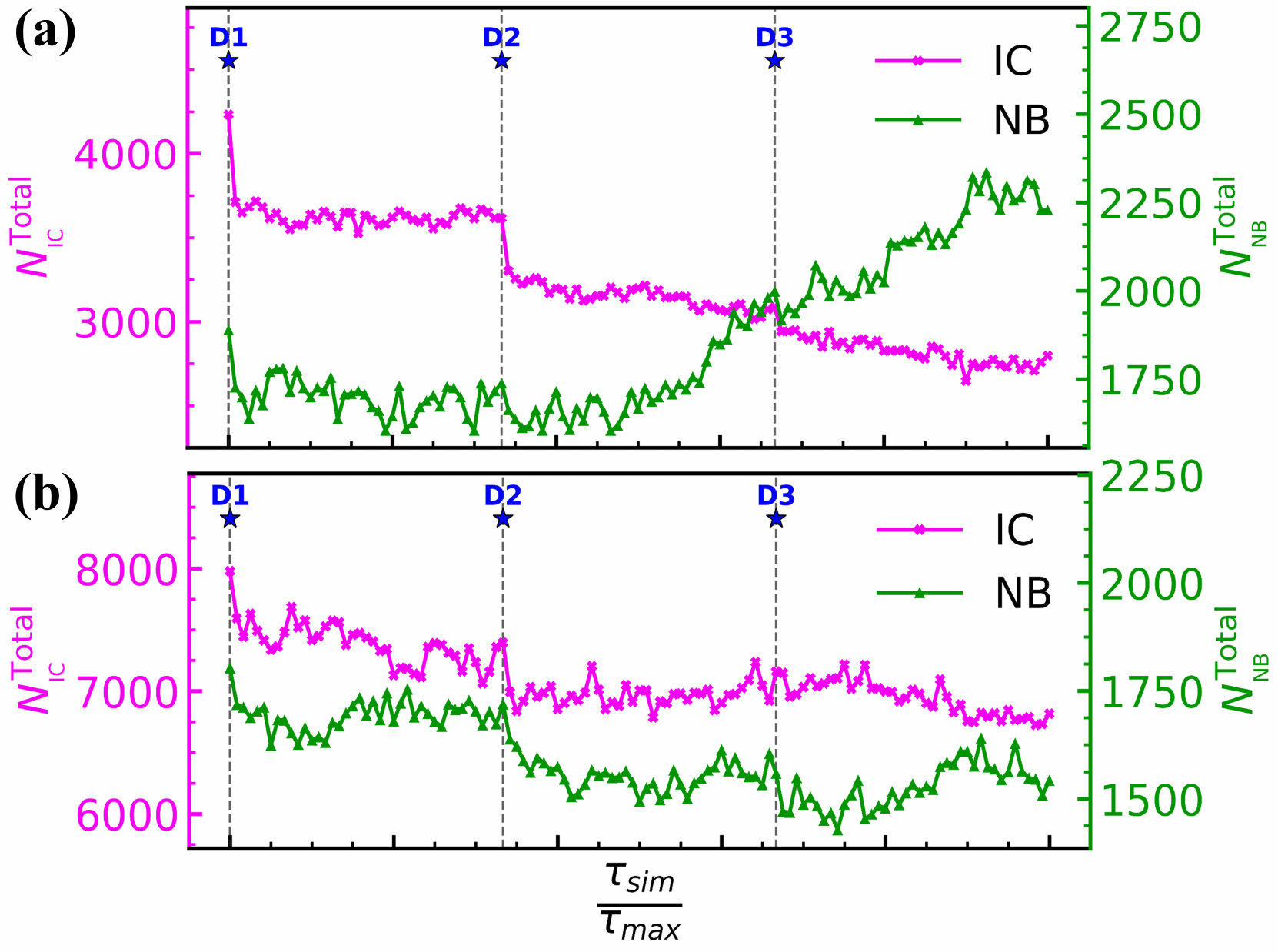}
    \caption{\small\textit{Activity-induced persistence of interchain contacts under faster depletion ($\lambda = 0.4$).} Time evolution of IC and NB contacts for (a) $\mathrm{Pe}=0$ and (b) $\mathrm{Pe}=1.0$ for $\rho_{_{M}} = 2.0$. Dashed vertical lines mark successive ABP depletion events D1-D3.}
    \label{fig:contdyn}
\end{figure}
\begin{figure}[h!]
    \centering
    \includegraphics[width=0.65\textwidth]{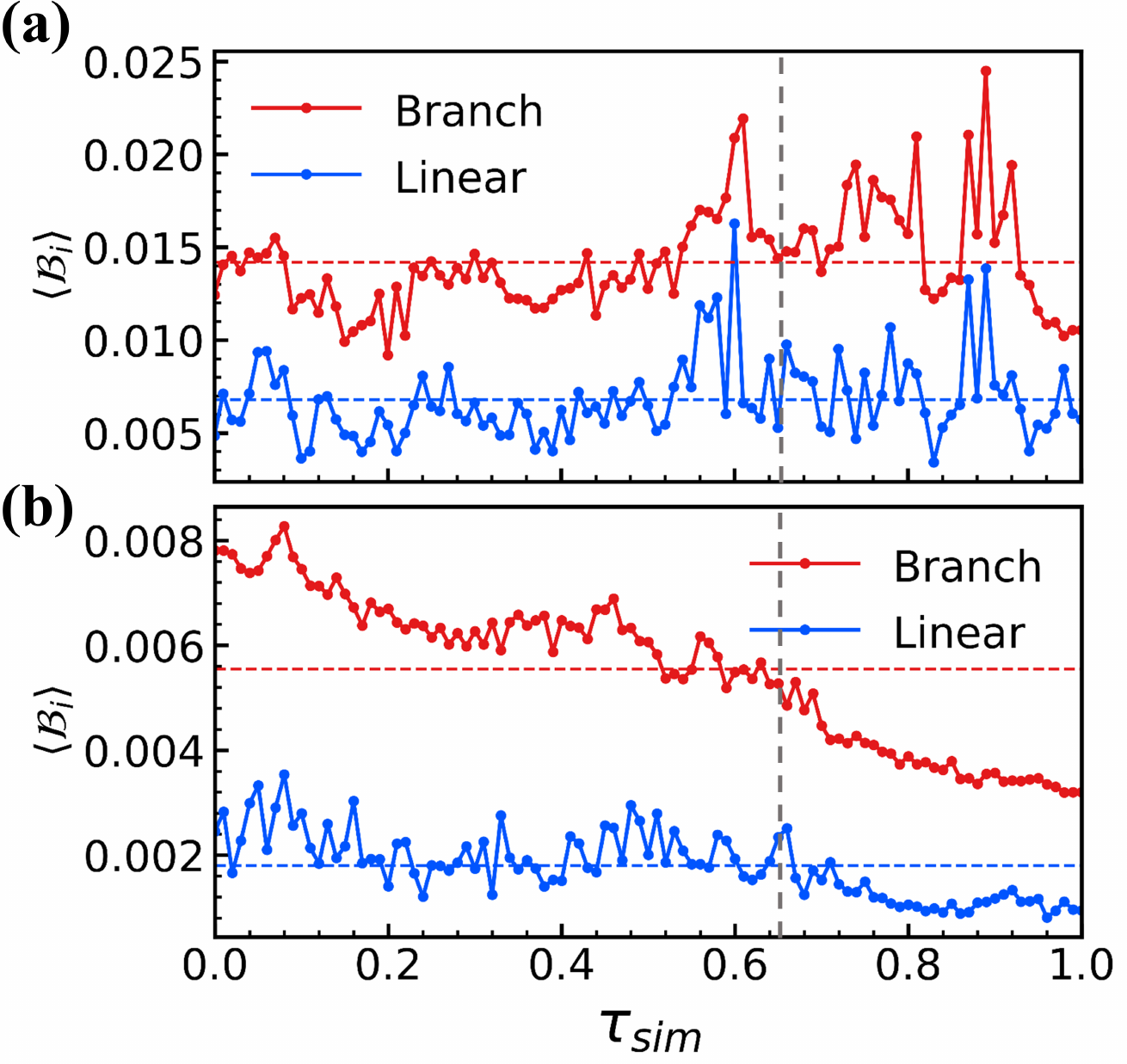}
    \caption{\textit{Activity suppresses bottleneck nodes and enhances network stability.} Time evolution of the mean betweenness centrality $\langle \mathcal{B}_i \rangle$ of branch and linear nodes for (a) $\text{Pe}=0$ and (b) $\text{Pe}=1.0$. Horizontal and vertical dashed lines mark time averages and the time at which all ABPs are removed, respectively.}
    \label{fig:stability}
\end{figure}
\begin{figure}[h!]
    \centering
    \includegraphics[width=0.95\textwidth]{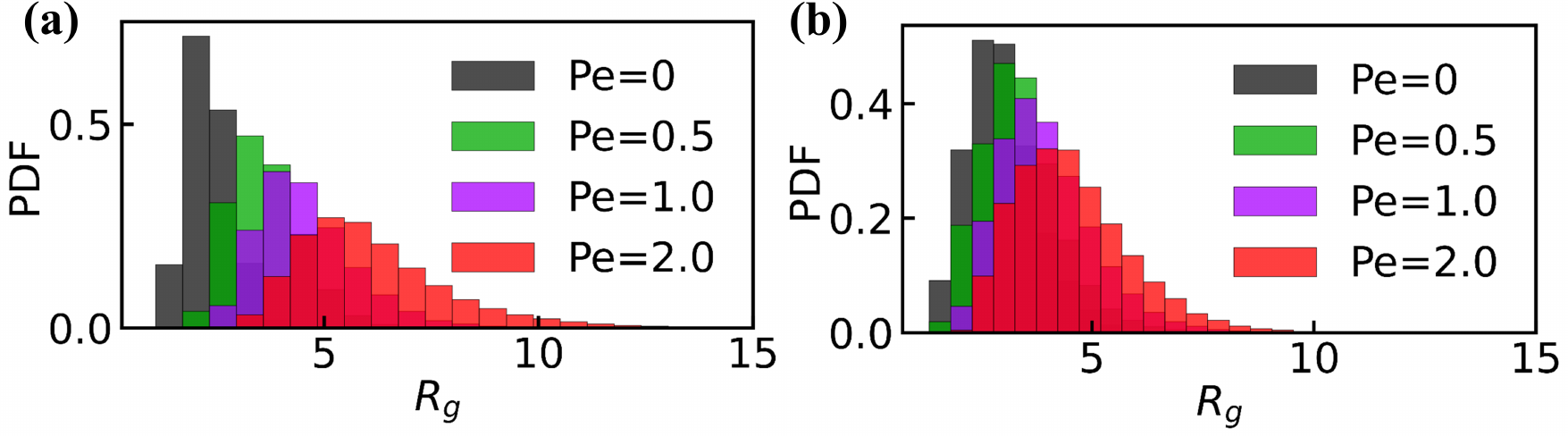}
    \caption{\small\textit{Polymer chains in the LCC (Largest connected component) network are more extended, promoting interchain connections.} 
    Probability distribution of $R_g$ for chains (a) in the LCC and (b) not in the LCC, across different $\mathrm{Pe}$ for $\rho_{_{M}} = 2.0$.}
    \label{fig:RgPDF}
\end{figure}
\clearpage
\subsection{Competing Length Scales Governing Activity-Induced Topological Imprinting}

\noindent The simulations presented here suggest that persistent network formation is not controlled by a single microscopic parameter but rather by the competition among several characteristic length scales. Although a quantitative theory remains to be developed, the present results motivate a scaling picture for the emergence of activity-induced topological imprinting.\\

\noindent The first microscopic length scale is the polymer monomer diameter, $\sigma_p$, which sets the local thickness of the polymer backbone. The second is the active Brownian particle (ABP) diameter, $\sigma_s$, whose ratio to the polymer diameter, $\sigma_s/\sigma_p$, controls the extent to which active particles can penetrate and reorganize the condensate. Our simulations indicate that very small ABPs produce little persistent restructuring, whereas sufficiently large ABPs promote polymer swelling (Fig.~\ref{fig:RgPDF}). However, when the ABPs become too large, steric constraints inhibit the close interchain encounters and winding events necessary to establish a system-spanning network. These observations suggest that network formation occurs only within an intermediate range of $\sigma_s/\sigma_p$.\\

\noindent A second important length scale is the polymer size itself, characterized by the radius of gyration, $R_g$. The radius of gyration determines the characteristic spatial extent over which an individual polymer can form bridges between neighboring regions of the condensate. If $R_g$ is much smaller than the characteristic network mesh size,
$\xi$, individual polymers cannot efficiently span neighboring pores and bridge formation is suppressed. Conversely, when $R_g$ becomes comparable to $\xi$, individual chains can simultaneously participate in multiple interchain contacts, facilitating the formation of a connected network backbone. For sufficiently large polymers, one may instead anticipate a crossover toward increasingly entangled or even kinetically arrested states. \\

\noindent A third length scale is the active persistence length,
\begin{equation}
\ell_a = \frac{v_0}{D_r},
\end{equation}
where $v_0$ is the propulsion speed and $D_r$ is the rotational diffusion coefficient of the active particles. Physically, $\ell_a$ measures the distance over which an active particle maintains a nearly persistent direction of motion before rotational diffusion randomizes its trajectory.\\

\noindent Now that the different lengthscales have been discussed, we postulate that persistent network formation is favored when the active persistence length becomes comparable to the characteristic internal dimensions of the condensate,
\begin{equation}
\ell_a \sim \xi \sim R_g.
\end{equation}
In this regime, active particles exert persistent forces over approximately one mesh spacing while polymer chains are sufficiently extended to bridge neighboring pores. Together, these conditions maximize the probability of repeated interchain encounters, chain wrapping, and the accumulation of topological constraints that stabilize the network.\\

\noindent Outside this regime, different behaviors are expected. When $\ell_a \ll \xi$, active particles behave primarily as local stochastic agitators whose motion resembles an enhancement of thermal fluctuations. Although local polymer conformations may fluctuate, active particles do not persistently displace chains over distances comparable to the network mesh size, making the formation of stable interchain bridges unlikely. In contrast, when
$\ell_a \gg \xi$, active particles travel coherently across many mesh spacings before changing direction. Rather than reinforcing local network connectivity, such long-persistence trajectories may instead promote large-scale restructuring of the condensate or continuously disrupt newly formed bridges before topological winding can accumulate. Likewise, when $R_g \ll \xi$, individual polymers are unable to connect neighboring regions of the network, whereas for $R_g \gg \xi$ the increasing density of entanglements may eventually drive the condensate toward a more arrested, solid-like state. \\

\noindent Collectively, these considerations suggest that activity-induced topological imprinting is governed by the competition among multiple characteristic length scales, including the polymer size, the active persistence length, the emergent mesh size, and the particle-size ratio $\sigma_s/\sigma_p$. Although the present work does not establish a predictive scaling theory, it provides a physical framework for future analytical descriptions of the transition between compact droplets, persistent percolated liquids, and ultimately mechanically arrested networks.
\clearpage
%\bibliography{Percolated_Condensates_Polymer}

\end{document}